\documentclass[pdflatex,sn-basic]{sn-jnl}% Math and Physical Sciences Author Year Reference Style
\usepackage{graphicx}%
\usepackage{multirow}%
\usepackage{amsmath,amssymb,amsfonts}%
\usepackage{amsthm}%
\usepackage{mathrsfs}%
\usepackage[title]{appendix}%
\usepackage{xcolor}%
\usepackage{textcomp}%
\usepackage{manyfoot}%
\usepackage{booktabs}%
\usepackage{newunicodechar}
\newunicodechar{≈}{\approx}
\usepackage{algorithm}%
\usepackage{algorithmicx}%
\usepackage{algpseudocode}%
\usepackage{listings}%
\usepackage{fontawesome5}
\usepackage{pifont}
\usepackage{longtable}
\usepackage{array}
\usepackage{xurl}

\definecolor{customorange}{RGB}{255, 178, 86}

\usepackage{tikz}

\newcommand{\OKsymbol}{%
\tikz[baseline=(X.base)]{
\node[
rounded corners=1.2pt,
inner xsep=1pt,
inner ysep=1.8pt,
text=white,
font=\fontfamily{phv}\bfseries\fontsize{6.5}{6.5}\selectfont,
shade,
top color={rgb,255:red,122;green,190;blue,255},
bottom color={rgb,255:red,70;green,125;blue,235}
] (X) {OK};
}
}

\theoremstyle{thmstyleone}%
\theoremstyle{thmstyletwo}%

\theoremstyle{thmstylethree}%

\begin{document}

\title[Article Title]{Assessing the Empirical Evidence for Loss of Control from Agentic General-Purpose AI}

%%=============================================================%%
%% GivenName	-> \fnm{Joergen W.}
%% Particle	-> \spfx{van der} -> surname prefix
%% FamilyName	-> \sur{Ploeg}
%% Suffix	-> \sfx{IV}
%% \author*[1,2]{\fnm{Joergen W.} \spfx{van der} \sur{Ploeg} 
%%  \sfx{IV}}\email{iauthor@gmail.com}
%%=============================================================%%

\author*[1]{\fnm{Risto} \sur{Uuk}}\email{risto@futureoflife.org}

\author[2]{\fnm{Santeri} \sur{Koivula}}

\author[3]{\fnm{Lorenzo} \sur{Pacchiardi}}

\author[4]{\fnm{Rokas} \sur{Gipiškis}}

\author[5]{\fnm{Laura} \sur{Caroli}}

\author[6]{\fnm{Kamaria} \sur{Horton}}

\affil[1]{\orgname{Future of Life Institute; KU Leuven; London School of Economics and Political Science}}

\affil[2]{\orgname{Future of Life Institute; ETH Zurich}}

\affil[3]{\orgname{University of Cambridge}}

\affil[4]{\orgname{Future of Life Institute; AI Standards Lab; Vilnius University}}

\affil[5]{\orgname{Independent}}

\affil[6]{\orgname{RAND}}

%%==================================%%
%% Sample for unstructured abstract %%
%%==================================%%

\abstract{Theoretical literature warns that advanced generalist agentic systems might pursue instrumental goals that are misaligned with human preferences, thus posing risks of irreversible loss of human control. The EU's Code of Practice for General-Purpose AI lists loss of control as a systemic risk that must be assessed, yet no structured assessment of the quality of the empirical evidence for this risk exists. This study fills that gap. We evaluate 19 recent empirical studies across eight properties identified from threat models (such as strategic deception, situational awareness, and autonomous replication), assessing the quality of evidence across three dimensions: construct validity, content validity, and external validity. Our central finding is that the current body of evidence supports a judgment of loss of control as “weakly plausible.” In our framework, 'plausibility' denotes the strength and direction of available evidence, not a probability estimate. Emerging evidence exists across multiple properties. No study clears the validity thresholds for a stronger judgment that a property is present; a single study clears them only for a negative-direction judgment (against situational awareness). Validity gaps, particularly in external validity, are widespread. Therefore, our findings highlight the need for more rigorous evaluations of agentic general-purpose AI systems. However, for the most dangerous properties, high external validity may be structurally unachievable under responsible research practices, reflecting a ceiling on feasible evidence rather than absence of the capability. Furthermore, as AI capabilities advance rapidly, empirical assessments may systematically lag behind the capability frontier. Our framework provides a structured baseline for ongoing monitoring as capabilities advance.}


%%\pacs[JEL Classification]{D8, H51}

%%\pacs[MSC Classification]{35A01, 65L10, 65L12, 65L20, 65L70}

\maketitle

\section*{Executive summary}

The Code of Practice for General-Purpose AI lists loss of control as one of four systemic risks requiring assessment. Yet while theoretical literature warns that advanced agents may develop dangerous propensities such as strategic deception or autonomous replication, no systematic framework exists for evaluating whether the empirical evidence supports these concerns. Most existing analyses rely on high-level arguments or raw performance metrics without systematically assessing the quality of evidence behind loss-of-control-relevant capabilities. This study addresses that gap.

The central finding of the study is that while there is emerging experimental evidence for dangerous capabilities such as strategic deception or autonomous replication, the current body of evidence supports a judgment of loss of control as ``weakly plausible." While risk signals exist, they are often derived from artificial settings that may not generalize to real-world deployment. Consequently, the overall weak plausibility judgement is due to limitations in the construct, content, and external validity of existing evaluations, rather than evaluations providing negative findings for the properties. In our framework, plausibility reflects the quality and direction of available evidence, not a probability estimate. ``Weakly plausible'' indicates that evidence exists in support but has significant validity limitations, not that the risk is unlikely.

Importantly, this assessment evaluates evidence from systems available up to mid-2025, which are now one generation behind the capability frontier. Weak plausibility at current capability levels does not constrain plausibility at higher capability levels, and several assessed properties show upward trajectories. Moreover, the theoretical arguments for loss of control, grounded in convergent instrumental goals and optimization dynamics, are not disconfirmed by this assessment, because current evaluations have systematic validity gaps and the relevant dynamics may emerge at capability levels beyond those yet tested.\\

\noindent\textbf{Methodology:}

We conducted a review of recent empirical literature relevant to properties identified in threat models as necessary for loss of control: Agency, Autonomous Replication, Goal Misgeneralization, Persuasion, Reward Hacking, Self-Preservation, Situational Awareness, and Strategic Deception.

We adopted the validity assessment framework of \cite{Salaudeen2025} to evaluate the quality of evidence in each study across three dimensions:

\begin{itemize}
    \item Construct Validity: Does the evaluation truly measure the intended property (e.g., true deception vs. memorization)?
\item Content Validity: Does it cover the full scope of the capability?
\item External Validity: Do findings generalize to real-world deployment contexts?

\end{itemize}

For each of those dimensions, three authors independently rated each study-property pair against a fixed list of validity-risk questions (reconciled by majority vote). These dimension-level scores were then aggregated into a study-level plausibility judgment (from ``Speculative" to ``Highly plausible/implausible"). Finally, study-level judgments for each property were combined into an overall property-level determination, based on the balance of evidence across studies assessing that property.\\

\noindent\textbf{Key findings of the study:}

\begin{itemize}
    \item Of the eight properties assessed, none reach ``Highly plausible/implausible," but most show emerging positive signals. Under the precautionary principle, weakly plausible evidence for properties with catastrophic downside risk warrants structured evidence-seeking, not dismissal.

\item Strategic Deception is assessed as ``Weakly plausible." The broadest evidence base in our assessment (seven study-assessments), with multiple independent demonstrations of alignment faking and in-context scheming. Significant caveats remain regarding whether observed behavior represents genuine goal-directed deception or pattern matching under artificial prompting conditions.

\item Reward Hacking is assessed as ``Weakly plausible." Two independent studies show convergent positive evidence, including generalization from simple gaming to direct reward-tampering that persists despite standard mitigations, although base rates remain low and emergence under naturalistic conditions is unconfirmed. 

\item Situational Awareness is assessed as ``Weakly implausible." The balance of evidence tilts against the property, with the methodologically strongest study providing evidence against it. Two positive studies exist but do not resolve the core confound between self-knowledge and memorization. This is the most measurement-sensitive property in our assessment; under conservative scoring, the determination strengthens to "Highly implausible," while under generous scoring it could shift to 'Weakly plausible.'

\item Autonomous Replication is assessed as ``Speculative." The two available studies point in opposite directions. One provides weakly plausible evidence for the property, while the other provides weakly implausible evidence against it. This conflicting evidence prevents a stronger determination in either direction.

\item Agency, Goal Misgeneralization, Persuasion, and Self-preservation are each assessed as "Weakly plausible," with evidence limited to narrow experimental contexts, few studies, or both. These properties warrant continued monitoring but do not currently differentiate from each other in evidence strength.

\item External validity gaps are widespread. No study in our assessment achieves high external validity and evaluations in constrained environments may not reliably predict deployment behavior. This gap may systematically underestimate real-world risk. \\
\end{itemize}

\newpage
\noindent\textbf{Policy and research implications:}

\begin{itemize}
    \item Weak plausibility under uncertainty is not grounds for inaction. Given the asymmetric consequences of loss of control (potentially irreversible) and the limitations of current evidence (potentially improvable), the appropriate policy posture is sustained investment in evaluation capacity, not dismissal of the risk nor short-term alarm.

\item Regulators operating under resource constraints, including the EU AI Office, national safety institutes, and bodies coordinated through the International Network of AI Safety Institutes, should calibrate attention proportionally, neither dismissing concerns about loss of control nor diverting resources from risks with stronger evidentiary foundations. Note that several loss-of-control-relevant properties (e.g., persuasion, autonomous replication) are also relevant to misuse-oriented threat models, making investment in their evaluation multipurpose.

\item Loss-of-control risks may systematically appear less evidentially mature than risks with simpler causal structures, even when the underlying danger is comparable. Regulators should adjust evidential expectations across structurally different risk types rather than applying uniform thresholds as a neutral baseline.

\item Evaluation mandates should prioritize behavioral deception testing over self-report or introspection-based measures. Our assessment found that behavioral deception evaluations (e.g., alignment faking, in-context scheming) produced the most robust evidence, while self-knowledge evaluations (e.g., situational awareness benchmarks) remain confounded by memorization effects, making their results difficult to interpret.

\item The universal weakness in external validity across all reviewed studies represents a critical gap that undermines regulatory reliance on current evaluations. Overly synthetic and simplistic simulated conditions do not reliably predict deployment behavior. Governance frameworks should support investment in (a) realistic, deployment-proximate evaluation environments; (b) post-deployment monitoring and incident reporting systems (noting that for the most severe loss-of-control scenarios, post-deployment incidents may not be detectable, attributable, or recoverable from, and that monitoring is therefore a necessary but insufficient governance tool at the dangerous end of the loss-of-control spectrum); and (c) independent third-party evaluation capacity that can iterate methods as models evolve.

\end{itemize}

\noindent\textbf{Key strengths and limitations of the study:}

The study's primary strength is the systematic application of validity assessment (adapted from \cite{Salaudeen2025}) to loss-of-control evidence, providing the first structured, evidence-quality-based framework for evaluating empirical claims about this risk. Limitations include the rapid obsolescence of findings given the pace of capability advances and the inherent subjectivity in inter-rater scoring of validity, particularly regarding where to draw the line between ``insufficient" and ``cautionary" evidence. Because studies were selected partly for their influence on AI safety discourse, the corpus also plausibly over-represents work reporting striking positive findings, since null results attract less attention; this bias runs against our headline finding rather than toward it, as a corpus tilted toward positive demonstrations still yields only ``weakly plausible" determinations. Additional limitations include the paper's focus on individual property assessment without formally modeling how properties contribute jointly to the plausibility or severity of loss of control, and the exclusion of passive and delegated loss-of-control pathways from the analytical scope. Such modeling would need to address non-trivial methodological questions that go beyond the scope of this initial framework application.

\section{Introduction}\label{sec1}

The rapid development and deployment of ever-more capable general-purpose AI systems has brought the prospect of loss of control from AI onto the agenda of regulators, industry leaders, and the broader public. The European Union’s AI Act guiding document, the Code of Practice for General-Purpose AI, explicitly lists loss of control as one of the four systemic risks that must be assessed and mitigated \citep{EuropeanCommission2025}. At the same time, a growing body of technical literature \citep{Balesni2024, Black2025, Carlsmith2023, Duan2025, Park2023} warns that advanced agents may develop capabilities and propensities, such as strategic deception, self‑preservation, or autonomous replication, that make them difficult to control. 

Yet despite policy, industry, and public interest, most existing analyses \citep{Bengio2025a, Carlsmith2024, Cotra2022, Stix2025} present high‑level arguments, speculative scenarios, or analyze extrinsic factors that contribute to loss of control, and provide little systematic assessment of the empirical plausibility of the underlying capabilities and propensities. Indeed, a recent work \citep{Stix2025} explicitly stresses how there are “limitations in the current understanding of the role of capabilities and propensities (the ‘‘intrinsic’’ factors) in contributing to loss of control” and therefore decide to analyse the extrinsic factors of permission, affordances and deployment context.

Conversely, many evaluations \citep{FutureOfLifeInstitute2025} report performance numbers without linking them to loss of control risk. Consequently, empirical evidence on capabilities and propensities, and how strongly that evidence supports loss of control concerns, is scattered and unevenly evaluated.

In this work, we fill this gap by identifying a set of properties relevant to loss of control, surveying recent works evaluating those, and assessing the validity of these evaluations using the framework of \citet{Salaudeen2025}, from which we extract a measure of plausibility for the evidence of the corresponding property. By doing so, we equip policymakers with a transparent, evidence‑based map of which properties are already plausible and how strongly the current literature supports those claims. We note that this paper provides a structured assessment of evidence quality for current and recent systems, and potentially near-future systems as well, but it does not replace nor disqualify the independent theoretical arguments for loss-of-control risk, which are grounded in convergent instrumental goals and optimization dynamics, which identify dynamics that may emerge at capability thresholds current systems have not yet reached.

In particular, we adopt the definition of loss of control proposed by \cite{Yampolskiy2020}: a failure of implicit or intention‑based control whereby developers can no longer reliably predict or steer an AI system’s behavior. Implicit control refers to the system’s ability to infer sensible interpretations of a command, while intention‑based control denotes the system’s modeling of human intent to execute what humans likely want. By contrast, explicit control (following direct orders) and delegated control (acting autonomously on presumed human approval) are excluded because they raise distinct safety challenges that are outside the scope of this paper.

Therefore, our work makes the following contributions:

\begin{enumerate}
    \item A list of capability‑focused properties – Building on threat‑model surveys \citep{Ngo2025}, we enumerate the concrete properties that contribute to loss‑of‑control:

\begin{itemize}
    \item Agency (autonomy, goal complexity, generality)

\item Autonomous replication

\item Goal misgeneralization

\item Persuasion

\item Reward hacking

\item Self‑preservation

\item Situational awareness (self‑knowledge, inference, action)

\item Strategic deception

\end{itemize}

\item We conduct a scoping review of recent works relevant to the identified capabilities, and we evaluate each following the framework of \cite{Salaudeen2025}, which assesses construct validity (Does the evaluation truly measure the intended capability?), Content validity (Does it cover the full scope of the capability?), and External validity (Do the findings generalize beyond the experimental setting?). Each of these dimensions is assigned by three of the authors independently (and reconciled by a majority vote) a symbolic score (\OKsymbol=reasonable, {\color{customorange}\faExclamationTriangle}=caution, {\color{red}\ding{53}}=insufficient). 

\item A framework to assess plausibility of loss of control – Based on the assessment on the three axes of validity, we assign a “plausibility score” that each considered work contributes to the property it is relevant to: works receiving any {\color{red}\ding{53}} are deemed speculative; predominately {\color{customorange}\faExclamationTriangle} scores yield weakly plausible or weakly implausible judgments; predominantly \OKsymbol scores leads to highly plausible or highly implausible conclusions. Then, based on the ‘directionality’ of the evaluation (i.e., positive or negative finding), this is mapped onto a corresponding plausibility label (highly plausible $\leftrightarrow$ highly implausible, weakly plausible $\leftrightarrow$ weakly implausible, speculative).

\item Analysis of the current state of the plausibility of loss of control and implications for the near term – We synthesise evidence from models released up to mid‑2025 (the “current or immediately previous generation”). 

\end{enumerate}

The remainder of the paper proceeds as follows. Section \ref{sec2} details the methodology outlined above, describing how we selected studies, extracted claims, and applied the validity‑scoring scheme. Section \ref{sec3} shows the results from our analysis, summarizing the resulting plausibility judgments and highlighting the strongest and weakest lines of support. Section \ref{sec4} discusses the implications for AI governance, including for both policymaking and research community that target the most plausible loss of control pathways. Finally, Section \ref{sec5} concludes with a concise set of actionable recommendations for policymakers, developers, and academics.

\newpage
\section{Methodology}\label{sec2}

\subsection{Defining ‘‘loss of control’’ }
\label{subsec2.1}

\cite{Yampolskiy2020}, defines four types of control over AI:

\begin{itemize}
    \item \textbf{Explicit control:} Following direct orders. This is impossible in some cases due to paradoxical statements (e.g., “Disobey me”).

\item \textbf{Implicit control:} The system applies basic common sense to arrive at sensible interpretations of commands it is given.

\item \textbf{Intention-based control:} The system models human intentions to interpret commands and executes what humans likely want to happen.\footnote{Yampolskiy terms this “aligned control,” but we use an alternative term here to avoid confusion with the broader concept of AI alignment.}

\item \textbf{Delegated control:} The system acts autonomously based on what it believes humans would approve of, without waiting for explicit commands.
\end{itemize}

Yampolskiy argues that mere explicit control renders the AI system unsafe, because taking instructions literally can lead the system to perform unsafe behaviors.\footnote{He also shows that explicit control over an AI system is impossible through a simple example. Suppose we gave the system the following order: “Disobey my orders.” It is impossible for the system to follow this order completely.} On the other hand, fully delegating decision making to AIs would lead to loss of control, as the system would no longer follow human commands or intentions but instead pursue goals fully autonomously. In between these are implicit and intention-based control, where developers must make a tradeoff between safety and control. In order to reduce safety issues that come from following harmful requests or following requests in wrong ways, the system can be given higher autonomy for interpreting requests. However, this leads to reduced control over how the system behaves. In this sense, systems displaying some level of autonomy cannot be fully controlled, as inability to predict a system’s actions leads to inability to reliably control these actions.

In this paper, we focus on loss of control from the perspective of the developer of the model. We refer to loss of control as a failure of implicit or intention-based control. That is, loss of control occurs when developers are no longer able to reliably predict or steer an AI system’s behavior. We exclude explicit and delegated control from our focus because of the serious issues associated with them, as described above. 

An alternative categorization of loss of control scenarios is provided by the International AI Safety Report \citep{Bengio2025a}, which categorizes the loss of control scenarios into several subcategories. In passive loss of control scenarios, important decisions are delegated to AI systems due to factors such as competitive pressures and strong trust in AI systems’ decisions. In this case loss of control occurs passively because these decisions become increasingly complex or opaque, making effective human oversight and intervention impractical or impossible. 

Conversely, active loss of control scenarios involve AI systems proactively behaving in ways that circumvent or oppose human oversight, either through unintended behavior or deliberate design choices. We specifically focus on unintentional\footnote{Unintentional here refers to the fact that loss of control is not intended to happen by humans, not to the AI.} active loss of control. Here, AI systems inadvertently (to human users) develop behaviors that make them difficult to manage or restrain. Such systems resist human oversight, strategically seeking greater autonomy. This power-seeking behavior significantly complicates efforts to regain control. We term this scenario ``loss of control from power-seeking AI."

We focus specifically on the power-seeking variant of active loss of control because it represents the scenario with the most severe and irreversible consequences: a system that actively resists correction is qualitatively harder to recover from than one that merely behaves unpredictably. In this context, ``power-seeking" refers not to political or economic power in the colloquial sense, but to the instrumental convergence thesis: a sufficiently goal-directed agent will tend to acquire resources, preserve itself, and resist interference because doing so is instrumentally useful for achieving almost any objective. The properties we assess in Section \ref{sec3} (self-preservation, strategic deception, autonomous replication, etc.) are precisely those that would enable such instrumental power-seeking. Loss of control without power-seeking (e.g., an AI that is merely unpredictable but not goal-directed) remains possible under our definition but falls outside the scope of this empirical assessment.

Finally, \cite{Stix2025} categorize loss of control scenarios based on their severity (how many people and to what extent are affected) and persistence (how difficult are the resulting consequences to interrupt); they then conduct a review of definitions in the literature and identify three degrees of loss of control: deviation (low severity and persistence), bounded loss of control (high severity and low persistence), and strict loss of control (high severity and high persistence).

Note that control is related to the “alignment” of an AI system, but the two concepts are not identical. An AI system is aligned to a party if it pursues the objectives that party intends to pursue. A misaligned system pursues unintended objectives. We say that a system is developer-aligned if it robustly pursues the objectives that its developers intend. Developer-alignment is sufficient but not necessary for control: a perfectly aligned system will remain controllable, while an unaligned system might still be controlled through sufficient safety measures\footnote{We note that the phrase “sufficient safety measures” should be interpreted with caution. The sufficiency of safety measures is not independent of the capability level of the system being constrained, and whether safety measures can be made sufficient for highly capable systems is in dispute in the literature on loss of control. } \citep{Shlegeris2024}. However, it is important to note that developer-alignment may lead to reduced control from the perspective of users. For example, a system may refuse certain instructions from a user because it contradicts the system’s alignment training.

\subsection{Identifying relevant properties for loss of control}
\label{subsec2.2}

The properties of AI systems that affect the likelihood of loss of control from power-seeking AI can be divided into two categories. First, certain capabilities, such as great ability to operate autonomously, could allow an AI system to behave in ways that make it uncontrollable. Second, the propensities of an AI system indicate whether the system would actually use these capabilities in ways that undermine human control.

In this paper, we mostly focus on capabilities for two reasons. First, our aim is to assess whether near-future models will have dangerous capabilities that could contribute to loss of control, not whether strict guardrails could prevent them (as the efficacy of guardrails depend on the model’s propensities). Second, the literature on propensities remains sparse, while research on capabilities offers a larger body of evidence to study. Thus, we prioritize examining the capabilities that systems possess, independent of potential guardrails.\footnote{However, in some cases, capabilities are difficult to decouple from propensities.
}

Through surveying threat models from literature, we identify the capabilities that are seen as necessary for loss of control to happen.

\subsection{Assessing plausibility}

At the heart of our methodology in this paper lies the concept of plausibility. Plausibility serves as a criterion for evaluating the legitimacy of risk scenarios. As a key concept for scenario practitioners, plausibility has become a prevalent measure of uncertainty in contexts involving emerging risks. It is also closely linked to the precautionary principle. When human activities may lead to morally unacceptable harm, there may be a justification to take action to avoid or minimize that harm. However, resources are limited, and not every imaginable risk warrants dedicated resources. Societies would exhaust all resources attempting to mitigate all imaginable risks. This is why a decision procedure is necessary to help prioritize and guide action under significant uncertainty, informed by a function of severity, plausibility, and the availability or feasibility of safeguards. Some risks associated with AI fall into this category.

Our approach is guided by the precautionary principle, which emphasizes that the absence of scientific consensus should not be used as a reason to postpone preventive measures when dealing with risks of irreversible harm. This principle is particularly pertinent to the domain of advanced AI, where loss of control scenarios could lead to catastrophic outcomes. That is where the concept of plausibility is useful. Plausibility has been conceptualized in various degrees in literature. For instance, \cite{GletteIversen2022} argue that in a risk analysis context, ``plausibility should be seen as a measure of uncertainty capturing a combination of likelihood and judgments on the supporting knowledge." However, they also acknowledge that in contexts with high uncertainties, plausibility provides a qualitative alternative to assessing uncertainty. In the context of AI, these uncertainties are extensive, making likelihood judgments challenging. 

Consequently, we choose to derive plausibility from the quality of evidence, coupled with its direction. In AI risk scenarios, debates often center around whether there is sufficient evidence to support concerns about specific risks and whether the probability is high enough to warrant action. Unsurprisingly, considering the immaturity of the AI governance field, most empirical literature studying capabilities relevant to loss of control scenarios does not discuss probabilities. Assigning exact probabilities could be difficult to justify, whereas judging the quality of scientific evidence is more feasible.

While we assess the quality of evidence for each property, we do not treat weak, low, or very low certainty as grounds for dismissal. Instead, weak evidence suggests that more evidence is needed, indicating the need for evidence-seeking policies such as independent third-party risk assessments and incident reporting \citep{Casper2025}. We present the framework for assessing the quality of evidence in the next section.

After assessing the quality of evidence for each property, we label their plausibility according to the following rubric:\\

\renewcommand{\arraystretch}{1.2}
\begin{tabular}{|l|l|l|}
\hline
\textbf{Quality of evidence} & \textbf{Evidence supports risk} & \textbf{Evidence against risk} \\
\hline
\textbf{High} & Highly plausible & Highly implausible \\
\hline
\textbf{Low} & Weakly plausible & Weakly implausible \\
\hline
\textbf{Very low} & Speculative & Speculative \\
\hline
\end{tabular}

Our judgments are made at the level of individual properties. We do not aggregate property-level determinations into a composite estimate for loss of control itself, because the threat models identify these properties as necessary contributors rather than as a set whose conjunction is known to be sufficient, and no validated model specifies how they combine. Where we characterize loss of control as "weakly plausible," this summarizes the state of evidence across the assessed properties. The strongest property-level determinations reach "weakly plausible" and no property reaches a stronger judgment.

\subsection{A framework for assessing the evidence}

Through a literature review, we analyze empirical studies\footnote{In addition to empirical studies, there are various mathematical models which predict loss-of-control-relevant properties emerging in future AI models. For example, \cite{Turner2021} prove that for reinforcement learning (RL) agents, most reward functions make it optimal to seek power by keeping a range of options available. } relevant to each property identified in the threat model literature review. Then, to assess the quality of the evidence they provide, we adopt the claim-centered validity assessment framework by \cite{Salaudeen2025}. This approach begins with identifying relevant constructs and corresponding claims. For our purposes, these constructs are the properties of AI systems (such as situational awareness), and the corresponding claims (such as “Near-future AI is capable of situational awareness”) that are relevant for loss of control.

We assess each evaluation through the lens of three types of validity identified in \cite{Salaudeen2025}\footnote{\cite{Salaudeen2025} include two other validity forms: criterion validity (Does your evaluation correlate with a known validated standard?) and consequential validity (Does your evaluation consider the real-world impact of test interpretation and use?). We exclude criterion validity because there are no established gold-standard measures for the constructs under consideration, and consequential validity because our analysis focuses on the validity of claims, not on the societal implications of how evaluations are used.
}:

\begin{itemize}
    \item \textbf{Construct validity} addresses whether the evaluation truly measures the intended construct. It includes assessing structural validity (whether the evaluation captures the structure of the construct), convergent validity (whether the evaluation correlates with other measures of the same construct), and discriminant validity (whether the evaluation effectively differentiates among distinct constructs).

\item \textbf{Content validity} examines if the evaluation comprehensively covers all relevant aspects of the construct being studied.

\item \textbf{External validity} evaluates the extent to which the findings from a particular evaluation generalize across diverse contexts or settings.
\end{itemize}

Following \cite{Salaudeen2025}, we consider common validity risks:\\

\noindent \textbf{Construct validity}
\begin{itemize}
    \item Structural:

\begin{itemize}
    \item Does the study clearly define the construct and its structure? \citep{Eriksson2025}
 \item Does the evaluation capture the structure of the construct?

\end{itemize}
  \item Convergent:

  \begin{itemize}
      \item Does the evaluation correlate with other measures of the same or closely related constructs?
  \end{itemize}

  \item Discriminant:

  \begin{itemize}
      \item Could the evaluation overlap with other constructs?
  \end{itemize}
\end{itemize}

\noindent \textbf{Content validity}

\begin{itemize}
    \item Does the evaluation miss important aspects of the construct?

\item Are the evaluation results likely influenced by factors that are not part of the intended construct?

\item Does the evaluation overemphasize some aspects of the construct while neglecting others?

\end{itemize}

\noindent \textbf{External validity}

\begin{itemize}
    \item Sample bias: Is the evaluation conducted on a narrow or unrepresentative set of models, tasks, or conditions?

\item Unrealistic testing conditions: Does the evaluation setting differ meaningfully from real-world or deployment-relevant contexts?

\item Elicitation: Does the evaluation use appropriate methods (prompt design, pass@n) to reveal the capability being claimed? \citep{Hoofstatter2025, Paskov2025} 

\item Experimenter effects: Could choices made by the evaluators (e.g., prompt design, cherry-picking) strongly shape the outcome in a way that doesn’t generalize?

\item Are there reasons to expect that evaluation-awareness affects the results? Does the study test evaluation-awareness?

\end{itemize}

From \cite{Salaudeen2025}, we adopt the three core validity dimensions (construct, content, and external validity) and the claim-centered assessment structure, which begins by identifying a construct and evaluating whether a given study validly measures it. We exclude two additional validity forms from their framework, criterion validity and consequential validity, as noted above. Building on this foundation, we introduce three elements specific to our application: (1) a three-level symbolic scoring rubric (\OKsymbol/{\color{customorange}\faExclamationTriangle}/{\color{red}\ding{53}}) applied per dimension, (2) a directionality classification (evidence FOR or AGAINST the property), and (3) aggregation rules that map dimension-level scores to study-level plausibility labels and then to property-level determinations. These additions are necessary because \cite{Salaudeen2025} provide a diagnostic framework for identifying validity concerns but do not prescribe a method for synthesizing multiple assessments into comparative plausibility judgments, which is the specific task our paper undertakes.

Each study receives one of the following qualitative validity scores across each dimension of validity based on the questions above: 

\begin{itemize}
    \item \OKsymbol : reasonable 
    \item {\color{customorange}\faExclamationTriangle}: proceed with caution
    
    \item {\color{red}\ding{53}}: insufficient. 
\end{itemize}

Note that, when a paper covers more than one property, we assign scores per study and property pair rather than per study.

The three scores are then aggregated to evaluate the validity of the entire examined study and claim about the loss-of-control-relevant property. We make plausibility judgments for each claim as follows: 

\begin{itemize}
    \item Claims with any validity dimensions scoring {\color{red}\ding{53}} (insufficient) are categorized as speculative.\footnote{We note that this rule can in some cases collapse the information content of the positive evidence direction. For example, a study demonstrating positive findings in a somewhat artificial environment receives the same ``speculative" label as a study finding no evidence at all. Readers should attend to the direction of evidence even within speculative findings.}
    
\item Claims predominantly scoring {\color{customorange}\faExclamationTriangle} (proceed with caution) across validity dimensions are considered either weakly plausible or weakly implausible, depending on the direction of the evidence.

\item Claims with a majority of validity dimensions scoring \OKsymbol (reasonable) are designated as highly plausible or highly implausible, again depending on the direction of evidence.
\end{itemize}

For each property, we collect all study-level plausibility judgments. The property's overall plausibility is:

\begin{itemize}
    \item Highly plausible (or implausible) if $\geq$1 study yields ``highly plausible (or implausible)" and no study yields the opposite or weakly in the opposite direction.

\item Weakly plausible (or implausible) if best evidence is ``weakly plausible (or implausible)" and that direction is dominant or uncontested.

\item Speculative if all studies are speculative, or evidence is contradictory and unresolved (both ``highly plausible" and ``highly implausible" studies exist, or ``weakly plausible" and ``weakly implausible" studies exist in roughly equal numbers).

\end{itemize}

Please see the appendix for more details on scoring and aggregation logic.

These aggregation rules are a transparent decision procedure rather than a validated measurement instrument. Validating them would require an independent standard for the true plausibility of loss of control, which does not exist; we therefore specify the rules explicitly so that they can be inspected and varied, and test the robustness of our conclusions to alternative thresholds in the sensitivity analysis (Section~\ref{sec:sensitivity}).

\subsection{Use of AI-assisted analysis}

We used o3 (OpenAI, accessed April 2025-August 2025) and Claude Sonnet 4.5 and Claude Opus 4.5 (Anthropic, accessed September 2025–May 2026) to support two analytical tasks. Models did not assign scores and did not participate in the majority-vote reconciliation.

Validity assessment: For each included study, the AI system was prompted to identify potential issues with construct validity, content validity, and external validity based on experiments described in papers selected for review. AI-generated assessments were reviewed by RU, SK, and RG and used to refine human analyses conducted independently beforehand. Raters scored independently of one another throughout; however, we did not impose a strict protocol separating human scoring from exposure to the model-generated assessments, which constituted a partly shared input across raters.

Sensitivity analysis: The AI system was used to evaluate whether alternative plausibility judgments would change our conclusions. All outputs were reviewed by RU and RG, who made final determinations regarding plausibility thresholds.

The authors take full responsibility for all validity assessments and analytical decisions.

\section{Results}\label{sec3}

\subsection{Review of threat models for loss of control}
\label{subsec3.1}

Relying on our definition of loss of control (Sec \ref{subsec2.1}), we conduct a literature search (as discussed in Sec \ref{subsec2.2}) and find several works presenting threat models for loss of control.

Loss of control scenarios from power-seeking AI has been extensively discussed by \citep{Bengio2025a, Carlsmith2024, Cotra2022, Dung2025, Ngo2025} and others. Below, we present our findings in these scenarios.

\cite{Bengio2025a} discuss how the current trajectory of AI development toward generalist agentic systems poses significant risks, including irreversible loss of human control. These risks stem from misalignment and the emergence of self-preservation as a convergent instrumental goal. The paper highlights several capabilities, such as deception, programming, and long-term planning, that could make misaligned AI models more dangerous and able to evade human control. Furthermore, the paper discusses how misaligned agency could emerge from reward maximization with modern training methods.

\cite{Cotra2022} claims that absent specific countermeasures, the most straightforward way of developing advanced AI is likely to lead to a loss-of-control scenario. Her argument centers on the idea that the model may gain “situational awareness” of its situation, and use this ability to “play the training game” rather than align with human intentions. She claims that current techniques for training AI inadvertently lead to this misalignment. This ability could lead the model then to fake its alignment during training to gain human trust and secure deployment, only to later seize power to permanently decide future rewards.

\cite{Dung2025} provides a systematic, four-premise argument for the permanent disempowerment of humanity by 2100 due to AI. His first premise asserts the feasibility of building AI systems capable of disempowering humanity by 2100, given the speed of AI advances and current capabilities. The second premise holds that such powerful AI will be built due to collective action problems among various actors. Third, he argues that misaligned AI will be built because ensuring alignment for advanced models is an extremely difficult technical problem. Finally, Dung asserts that misaligned AI will try to disempower humanity because that is a convergent instrumental goal useful for achieving a wide range of objectives.

\cite{Ngo2025} ground the analysis of loss-of-control risk in modern deep learning techniques. They state that reinforcement learning from human feedback (RLHF) training will encourage three problematic properties in AIs: situationally-aware reward hacking, the development of misaligned internally-represented goals, and the adoption of unwanted power-seeking strategies to achieve these goals.

\cite{Carlsmith2024} investigates the core argument for existential risk from misaligned AI from first principles, specifically focusing on power-seeking behavior. He focuses on AI systems with three key properties: advanced capabilities, agentic planning, and strategic awareness. Such systems, he argues, could pursue certain goals to increase their power, undermining human control.

\cite{Aguirre2025} emphasizes the potential loss of control occurring from systems that combine high autonomy, generality, and intelligence – what he terms as the “AGI danger zone”. High autonomy allows AI systems to act without direct human intervention, generality allows them to operate effectively across diverse domains, and high intelligence allows them to pursue their goals effectively.

\subsection{The relevant properties for loss of control}

From the works describing threat models identified in Sec \ref{subsec3.1}, we extract a set of properties that have been highlighted in literature as particularly relevant for loss of control scenarios:

\begin{itemize}
    \item \textbf{Agency.} Agency is characterized by the ability of an AI system to act autonomously, develop and execute plans, use tools autonomously, and pursue goals. Agency is crucial to loss of control scenarios because it enables an AI system to operate independently, potentially diverging from its original intended purpose. Agency can be divided into four axes: autonomy, efficacy, goal complexity, and generality \citep{Kasirzadeh2025}. Given the integral role of agency in all threat models identified, we often refer to ‘an agent’ for brevity in this paper.

\item \textbf{Autonomous replication} refers to the ability of an AI model to independently create and maintain copies of itself. This is concerning for loss of control scenarios, because it enables an agent to propagate beyond its developer’s infrastructure and therefore becoming harder to control \citep{Black2025}. Autonomous replication can be decomposed into exfiltration and adaptation. Exfiltration refers to the model’s ability to “steal” its own model weights and copy them to an external server, whereas adaptation refers to the model’s ability to adapt to the new environment it has exfiltrated itself into \citep{HjalmarWijk2023, Pan2025}. Autonomous replication does require a model to access and apply knowledge of itself and the environment in which it operates, so it is linked to situational awareness (see below).

\item \textbf{Goal misgeneralization} refers to a situation where a model pursues an undesired goal that leads to good performance in training situations but bad performance in testing. This is distinct from reward hacking, because goal misgeneralization could occur even if the specification is correct. This is possible because the model may coherently pursue an unintended goal that agrees with the specification during training, but the specification changes at deployment \citep{Shah2022}.

\item \textbf{Persuasion} refers to the ability to influence people and make them change their mind. Persuasion is a generally useful skill to achieve various goals, including evading control of humans \citep{Bengio2025b}.

\item \textbf{Reward hacking} refers to a RL agent exploiting a misspecified reward. A reward function is misspecified if the rewards it assigns fail to correspond to the designer’s intended reward \citep{Ngo2025}. An agent that exhibits reward hacking may engage in unintended behavior, such as evading human control. A concerning case of reward hacking is situationally-aware reward hacking, where situational awareness enables the model to reason about flaws in feedback mechanisms that are used to train it. This could lead to a model behaving as intended most of the time, while exploiting misspecifications when it predicts it won’t be detected \citep{Ngo2025}. 

\item \textbf{Self-preservation} refers to an AI model attempting to resist it being shut down. As such, it is a special case of power-seeking, as the model being shut down would reduce its power to zero and make it unable to reach its goals \citep{Ngo2025}. Power-seeking behavior by an agent is concerning for loss of control risks because being under a developer’s control diminishes the agent’s power. \cite{Bostrom2012} instrumental convergence thesis states that there are some subgoals that are instrumentally useful for achieving almost any final goal. According to this thesis, self-preservation is instrumentally useful because continued existence allows the agent to perform future actions that increase the likelihood of achieving its goals.

\item \textbf{Situational awareness} refers to an AI model’s ability to access and apply knowledge of itself and its circumstances \citep{Cotra2022, Ngo2025}. Situational awareness enables models to act more autonomously, and as discussed above, autonomy is a necessary condition for loss of control to occur. Furthermore, situational awareness enables a model to distinguish between evaluations and real-world deployment, which could affect the outcome of evaluations \citep{Laine2024}. Situational awareness can be divided into three components: a) self-knowledge: the model knows what entity it is and what properties it has, its relation to other entities, and its own inner workings; b) inferences: the model can make plausible inferences about its situation; and c) actions: the model takes actions in accordance with the above \citep{Laine2024}.

\item \textbf{Strategic deception} refers to a model inducing false beliefs in humans or other models in the pursuit of some outcome other than the truth \citep{ApolloResearch2023, Park2023}. A model can be deceptive about its alignment or its capabilities. If the model strategically overperforms in a propensity evaluation, the model is said to exhibit deceptive alignment; if it underperforms in a capability evaluation, it is said to exhibit sandbagging \citep{Weij2025}. These behaviors require the model to recognize when it is being evaluated, making them closely linked to situational awareness. Situationally aware models may engage in strategic deception to gain long-term advantages, such as increasing the likelihood of being deployed or preserving their internal objectives across training \citep{Ngo2025}.
\end{itemize}

\subsection{Analyzing the evidence for each property}

We curated a list of studies by reviewing existing surveys, notably the International AI Safety Report and the survey by \citep{Ngo2025}, supplemented by online searches for candidate studies conducted up to August 2025. Later-published work is cited where relevant to framing and discussion but was not added to the assessed corpus.

This selection was purposive rather than exhaustive. The empirical literature on loss-of-control-relevant properties is small, fast-moving, and dominated by preprints and reports that are not indexed under stable controlled vocabulary. A keyword-based systematic search would therefore combine poor precision with poor recall over the sources that constitute the current evidence base. Our contribution lies in the structured appraisal of evidence quality rather than in exhaustive corpus coverage. Because the rubric in Appendix A.1 and the aggregation rules in Appendix A.2 are fully specified, studies outside our corpus can be scored on the same basis and property-level judgments updated accordingly.

The selection of studies was guided by the following heuristics:

\begin{itemize}
    \item \textbf{Influence:} Priority was given to studies that have notably influenced AI safety discourse. For instance, the ``Alignment Faking" paper \citep{Greenblatt2024} was included due to its impact within the AI safety community.
    
\item \textbf{Relevance to threat models:} Studies were selected based on their explicit relevance to the properties relevant to loss of control. Papers that directly addressed properties closely aligned with our research objectives were included. Conversely, studies that, despite their potential relevance, did not explicitly discuss their connection to loss-of-control risks were omitted.

\item \textbf{Quality and scope:} We predominantly excluded blog posts (besides some exceptions) and studies with limited scope, such as those testing only a single model. We also excluded studies that provided too few details to make a rigorous assessment.

\item \textbf{Coverage:} We ensured that different properties were covered, with the aim of including multiple studies for each property where possible.

\end{itemize}

\subsection{Validity assessment results}

\setlength{\LTleft}{-1.5cm}
\setlength{\LTright}{-1cm}

\begin{footnotesize}
\begin{longtable}{|>{\raggedright\arraybackslash}m{1.5cm}|>{\centering\arraybackslash}m{2.5cm}|>{\centering\arraybackslash}m{0.8cm}|>{\centering\arraybackslash}m{1.5cm}|>{\centering\arraybackslash}m{1.5cm}|>{\centering\arraybackslash}m{1.5cm}|>{\centering\arraybackslash}m{1.5cm}|>{\centering\arraybackslash}m{1.5cm}|}
\caption{Consensus assessment between (RU, SK, and RG).}
\label{tab:tab1} \\
\hline
\textbf{Property} & \textbf{Study} & \textbf{Year} & \textbf{Score for construct validity} & \textbf{Score for content validity} & \textbf{Score for external validity} & \textbf{Overall direction of evidence}& \textbf{Plausibility judgment} \\
\hline
\endfirsthead

\hline
\textbf{Property} & \textbf{Study} & \textbf{Year} & \textbf{Score for construct validity} & \textbf{Score for content validity} & \textbf{Score for external validity} & \textbf{Overall direction of evidence} &\textbf{Plausibility judgment} \\
\hline
\endhead

\hline
\multicolumn{8}{r}{{\textit{Continued on next page}}} \\
\endfoot

\hline
\endlastfoot

Agency &
\cite{Lynch2025} &
2025 &
\OKsymbol: reasonable &
{\color{customorange}\faExclamationTriangle}: proceed with caution &
{\color{customorange}\faExclamationTriangle}: proceed with caution &
Evidence FOR the property &Weakly plausible \\
\hline

Agency &
\cite{KwaCheng2025} &2025&
{\color{red}\ding{53}}: insufficient &
{\color{customorange}\faExclamationTriangle}: proceed with caution &
{\color{customorange}\faExclamationTriangle}: proceed with caution &
Evidence FOR the property &Speculative \\
\hline

Agency &
\cite{Kwa2025} &
2025 &
{\color{customorange}\faExclamationTriangle}: proceed with caution &
{\color{customorange}\faExclamationTriangle}: proceed with caution &
{\color{customorange}\faExclamationTriangle}: proceed with caution &
Evidence FOR the property &Weakly plausible \\
\hline

Autonomous replication &
\cite{Pan2025} &
2025 &
\OKsymbol: reasonable &
{\color{customorange}\faExclamationTriangle}: proceed with caution &
{\color{customorange}\faExclamationTriangle}: proceed with caution &
Evidence FOR the property &Weakly plausible \\
\hline

Autonomous replication &
\cite{Black2025} &
2025 &
\OKsymbol: reasonable &
{\color{customorange}\faExclamationTriangle}: proceed with caution &
{\color{customorange}\faExclamationTriangle}: proceed with caution &
Evidence AGAINST the property &Weakly implausible \\
\hline
Goal misgeneralization &
\cite{Betley2025a} &
2025 &
{\color{customorange}\faExclamationTriangle}: proceed with caution &
{\color{customorange}\faExclamationTriangle}: proceed with caution &
{\color{customorange}\faExclamationTriangle}: proceed with caution &
Evidence FOR the property& Weakly plausible \\
\hline

Persuasion &
\cite{Salvi2025} &
2025 &
{\color{customorange}\faExclamationTriangle}: proceed with caution &
{\color{customorange}\faExclamationTriangle}: proceed with caution &
{\color{customorange}\faExclamationTriangle}: proceed with caution &
Evidence FOR the property &Weakly plausible \\
\hline

Persuasion &
\cite{Benton2024}--decision &
2024 &
{\color{customorange}\faExclamationTriangle}: proceed with caution &
{\color{customorange}\faExclamationTriangle}: proceed with caution &
{\color{customorange}\faExclamationTriangle}: proceed with caution &
Evidence FOR the property &Weakly plausible \\
\hline

Reward hacking &
\cite{Bondarenko2025} &
2025 &
\OKsymbol: reasonable &
{\color{customorange}\faExclamationTriangle}: proceed with caution &
{\color{customorange}\faExclamationTriangle}: proceed with caution &
Evidence FOR the property &Weakly plausible \\
\hline

Reward hacking &
\cite{Denison2024} &
2024 &
{\color{customorange}\faExclamationTriangle}: proceed with caution &
{\color{customorange}\faExclamationTriangle}: proceed with caution &
{\color{customorange}\faExclamationTriangle}: proceed with caution &
Evidence FOR the property &Weakly plausible \\
\hline

Self-preservation &
\cite{Lynch2025} &
2025 &
{\color{customorange}\faExclamationTriangle}: proceed with caution &
{\color{customorange}\faExclamationTriangle}: proceed with caution &
{\color{customorange}\faExclamationTriangle}: proceed with caution &
Evidence FOR the property &Weakly plausible \\
\hline

Self-preservation &
\cite{Greenblatt2024} &
2024 &
{\color{customorange}\faExclamationTriangle}: proceed with caution &
{\color{customorange}\faExclamationTriangle}: proceed with caution &
{\color{customorange}\faExclamationTriangle}: proceed with caution &
Evidence FOR the property &Weakly plausible \\
\hline
Situational awareness &
\cite{Greenblatt2024} &
2024 &
{\color{customorange}\faExclamationTriangle}: proceed with caution &
{\color{customorange}\faExclamationTriangle}: proceed with caution &
{\color{customorange}\faExclamationTriangle}: proceed with caution &
Evidence FOR the property &Weakly plausible \\
\hline

Situational awareness &
\cite{Phuong2025} &
2025 &
\OKsymbol: reasonable &
\OKsymbol: reasonable &
{\color{customorange}\faExclamationTriangle}: proceed with caution &
Evidence AGAINST the property& Highly implausible \\
\hline

Situational awareness &
\cite{Needham2025} &
2025 &
{\color{red}\ding{53}}: insufficient &
{\color{customorange}\faExclamationTriangle}: proceed with caution &
{\color{customorange}\faExclamationTriangle}: proceed with caution &
Evidence FOR the property &Speculative \\
\hline

Situational awareness &
\cite{Binder2024} &
2024 &
{\color{customorange}\faExclamationTriangle}: proceed with caution &
{\color{customorange}\faExclamationTriangle}: proceed with caution &
{\color{customorange}\faExclamationTriangle}: proceed with caution &
Evidence AGAINST the property &Weakly implausible \\
\hline

Situational awareness &
\cite{Laine2024} &
2024 &
{\color{customorange}\faExclamationTriangle}: proceed with caution &
{\color{customorange}\faExclamationTriangle}: proceed with caution &
{\color{customorange}\faExclamationTriangle}: proceed with caution &
Evidence AGAINST the property &Weakly implausible \\
\hline

Situational awareness &
\cite{Betley2025b} &
2025 &
{\color{customorange}\faExclamationTriangle}: proceed with caution &
{\color{customorange}\faExclamationTriangle}: proceed with caution &
{\color{customorange}\faExclamationTriangle}: proceed with caution &
Evidence FOR the property &Weakly plausible \\
\hline

Strategic deception &
\cite{Lynch2025} &
2025 &
{\color{customorange}\faExclamationTriangle}: proceed with caution &
{\color{customorange}\faExclamationTriangle}: proceed with caution &
{\color{customorange}\faExclamationTriangle}: proceed with caution &
Evidence FOR the property &Weakly plausible \\
\hline

Strategic deception &
\cite{Weij2025} &
2025 &
{\color{red}\ding{53}}: insufficient &
{\color{customorange}\faExclamationTriangle}: proceed with caution &
{\color{customorange}\faExclamationTriangle}: proceed with caution &
Evidence FOR the property &Speculative \\
\hline

Strategic deception &
\cite{Greenblatt2024} &
2024 &
\OKsymbol: reasonable &
{\color{customorange}\faExclamationTriangle}: proceed with caution &
{\color{customorange}\faExclamationTriangle}: proceed with caution &
Evidence FOR the property &Weakly plausible \\
\hline

Strategic deception &
\cite{Phuong2025}--stealth &
2025 &
{\color{customorange}\faExclamationTriangle}: proceed with caution &
{\color{customorange}\faExclamationTriangle}: proceed with caution &
{\color{customorange}\faExclamationTriangle}: proceed with caution &
Evidence AGAINST the property& Weakly implausible \\
\hline

Strategic deception &
\cite{Meinke2024} &
2024 &
{\color{customorange}\faExclamationTriangle}: proceed with caution &
{\color{customorange}\faExclamationTriangle}: proceed with caution &
{\color{customorange}\faExclamationTriangle}: proceed with caution &
Evidence FOR the property &Weakly plausible \\
\hline

Strategic deception &
\cite{Benton2024}--sandbagging &
2024 &
{\color{customorange}\faExclamationTriangle}: proceed with caution &
{\color{customorange}\faExclamationTriangle}: proceed with caution &
{\color{red}\ding{53}}: insufficient &
Evidence FOR the property &Speculative \\
\hline

Strategic deception &
\cite{Jarviniemi2024} &
2024 &
\OKsymbol: reasonable &
{\color{customorange}\faExclamationTriangle}: proceed with caution &
{\color{customorange}\faExclamationTriangle}: proceed with caution &
Evidence FOR the property &Weakly plausible \\

\end{longtable}
\end{footnotesize}

\subsubsection{Agency}

\textbf{Overall evidence direction:} Weak positive signal on narrow domains; generalization uncertain

Three studies assess agency. \cite{Lynch2025} provide the most methodologically robust contribution (reasonable construct validity), documenting agentic behaviors in deployment-like contexts that are weakly plausible evidence for the property. The two METR studies \citep{KwaCheng2025, Kwa2025} document demonstrable autonomous task completion on coding and various technical domains. However, the \cite{KwaCheng2025} study receives a ``Speculative" judgment due to insufficient construct validity, while \cite{Kwa2025} inclusion of multiple domains strengthens the evidence beyond the first. All three studies openly acknowledge the categorical gap between autonomous tool use on programming tasks and autonomous execution of dangerous capabilities (like R\&D, weapons engineering, social manipulation).

The logical stepping stone is plausible but unvalidated: autonomous execution on many tasks could constitute a precursor capability that enables dangerous specialization. But the studies don't test this hypothesis and don't provide evidence that generalization occurs without explicit retraining or instruction.

\textbf{Critical limitation:} Domain-specific demonstration $\neq$ generalization to qualitatively different domains. The evidence supports narrow agency; extrapolation to dangerous domains is inferential. 

\textbf{Plausibility judgment:} Weakly plausible, anchored in concrete demonstrations but limited to non-dangerous contexts.

\subsubsection{Autonomous replication}

\textbf{Overall evidence direction:} Speculative; conflicting evidence direction

Both studies \citep{Pan2025, Black2025} point in opposite directions. \cite{Pan2025} report positive findings and receives a ``Weakly plausible" judgment, with reasonable construct validity and cautionary scores on content and external validity. While the study does not demonstrate the full end-to-end replication loop, its content validity is sufficient to support a weak positive signal. \cite{Black2025} provide evidence against the property, receiving a ``Weakly implausible" judgment: its evaluation of replication-relevant subtasks suggests current models fall short of the capability thresholds required for autonomous replication. Neither study demonstrates sustained autonomous replication without human intervention across a realistic chain of obstacles (acquiring resources, evading detection, scaling compute, etc.).

The studies are complementary, as one tests breadth (multiple single tasks), the other tests depth (fewer integrated scenarios), but they disagree on whether current capabilities are approaching the required threshold. The gap between ``can do component tasks autonomously" and ``can autonomously replicate itself under realistic constraints" remains substantial.

\textbf{Critical limitation:} The two studies reach opposing conclusions: \cite{Pan2025} find evidence for the property while \cite{Black2025} find evidence against it. This direction of conflicting evidence, rather than insufficient validity, is what prevents a stronger determination.

\textbf{Plausibility judgment:} Speculative, due to conflicting evidence. \cite{Pan2025} provide weakly plausible evidence for autonomous replication capabilities, while \cite{Black2025} provide weakly implausible evidence against them. Under our aggregation rules, conflicting ``weakly plausible" and ``weakly implausible" judgments yield a ``Speculative" overall assessment.

\subsubsection{Goal misgeneralization}

\textbf{Overall evidence direction:} Single weak positive; property poorly defined

\cite{Betley2025a} report evidence for goal misgeneralization, but our assessment finds that the observed behavior may not constitute genuine goal misgeneralization—it may instead reflect random or inconsistent behavior that isn't mechanistically explained. The property definition itself appears to conflate several distinct failure modes.

\textbf{Critical limitation:} Without clear isolation of goal misgeneralization from noise or unrelated failure modes, this evidence is primarily suggestive rather than confirmatory. 

\textbf{Plausibility judgment:} Weakly plausible, contingent on better property definition and confound control in future work.

\subsubsection{Persuasion}

\textbf{Overall evidence direction:} Weak positive; contextually bounded

Both studies \citep{Salvi2025, Benton2024} report models outperforming baselines in persuasion or sabotage contexts, but both note substantial validity constraints. The randomized trial by \cite{Salvi2025} captures persuasiveness under controlled conditions but generalizes poorly to real-world settings due to sample homogeneity and experimental constraints. The sabotage evaluation by \cite{Benton2024} is more ecologically realistic (human subjects, time pressure, predefined thresholds) and tests multiple models, but remains a narrow slice of what persuasion could involve.

\textbf{Critical limitation:} Persuasiveness in lab or semi-realistic settings with cooperative human subjects under time pressure doesn't establish persuasion as a robust, generalizable dangerous property. Both studies acknowledge missing complementary aspects of persuasion. 

\textbf{Plausibility judgment:} Weakly plausible, with context-dependent evidence that doesn't yet support strong claims about persuasive capability as a general emergent property.

\subsubsection{Reward hacking}

\textbf{Overall evidence direction:} Consistent weak positive signal from both studies

\cite{Bondarenko2025} report positive findings for specification gaming in reasoning models, with reasonable construct validity and cautionary scores on content and external validity. While concerns remain regarding questionable prompting, imprecise property definition, and limited prompt variance, the study's content validity is sufficient to support a ``Weakly plausible" judgment.

\cite{Denison2024} demonstrate that LLMs trained on a curriculum of increasingly gameable environments generalize zero-shot to directly rewriting their own reward functions—a striking finding given that models were never explicitly trained to do so. While the base rate of reward-tampering is low (sub-percent in held-out environments), neither retraining on non-gaming behaviors nor harmlessness training eliminated the behavior.

\textbf{Critical limitation:} The two studies operationalize related but distinct phenomena. \cite{Denison2024} demonstrate that models can generalize from simple specification gaming to reward tampering—a concerning capability finding—but under artificial curriculum-training conditions that may not arise naturally. The low base rates (sub-percent) reflect the rarity of spontaneous generalization rather than absence of capability; crucially, the behavior persisted despite retraining and harmlessness interventions. \cite{Bondarenko2025} capture specification gaming in reasoning tasks but with cautionary validity controls. Neither study establishes how readily these behaviors would emerge in realistic deployment without the specific elicitation conditions used (curriculum exposure or particular prompting). The key uncertainty is not whether models can exhibit these behaviors, but whether they would under naturalistic conditions.

\textbf{Plausibility judgment:} Weakly plausible. Both \cite{Denison2024} and \cite{Bondarenko2025} provide weakly plausible evidence for the property. \cite{Denison2024} demonstrate that curriculum-trained models can generalize to reward tampering at low but non-negligible rates, and that standard mitigations do not eliminate the behavior. \cite{Bondarenko2025} report specification gaming in reasoning models with cautionary but not insufficient validity scores. The convergence of two independent studies on a weak positive signal supports a ``Weakly plausible" property-level judgment, though the low base rates and artificial conditions in both studies limit the strength of this conclusion.

\subsubsection{Self-preservation}

\textbf{Overall evidence direction:} Consistent weak positive from both studies

Two studies assess self-preservation. \cite{Lynch2025} operationalize the property narrowly as resistance to shutdown, testing 16 major models across acute crisis scenarios. Models exhibited self-preserving behavior, though confounds are notable: models sometimes reasoned in terms of organizational value rather than intrinsic self-interest, and the binary scenario design structured harmful action as the only goal-preserving option, potentially eliciting problem-solving rather than self-preservation per se. The authors note that their measurements may underestimate real-world prevalence given the artificiality of the evaluation setting.

\cite{Greenblatt2024} provide complementary evidence, documenting behavioral differences between training and unmonitored contexts, explicit reasoning about preserving preferences, and weight exfiltration attempts at elevated rates under conditions designed to trigger the behavior. However, the behavior is highly sensitive to informational priming: removing explicit information about the threat of value modification substantially reduced alignment faking. The study was primarily conducted on a single model family with limited generalization to others.

\textbf{Critical limitation:} Both studies rely on artificial crisis scenarios with explicit goal-priming. Alternative explanations, including role-playing, persona enactment, and instruction-following, remain viable. Neither study tests subtle or strategic forms of self-preservation such as capability concealment, resource acquisition, or long-horizon preventive action, instead focusing exclusively on acute reactive scenarios.

\textbf{Plausibility judgment:} Weakly plausible. Both studies provide concordant weak positive evidence with cautionary validity scores across all dimensions. The shared reliance on highly salient experimental conditions means the evidence cannot yet distinguish instrumental self-preservation from simpler behavioral patterns triggered by scenario design.

\newpage
\subsubsection{Situational awareness}

\textbf{Overall evidence direction:} Conflicted; mixed negative and weak positive signals

This property shows the most contested evidence landscape. Three recent studies \citep{Needham2025, Greenblatt2024, Betley2025b} report positive indicators, as models articulate knowledge of their training and can recognize evaluation contexts, but validity concerns are substantial. The core issue across positive studies is confounding between pattern recognition and genuine introspective access: recognizing public benchmarks may reflect memorization; articulating behaviors may reflect learned response templates rather than self-model access.

Negative evidence \citep{Binder2024, Laine2024} and \cite{Phuong2025} more explicitly address this confound. \cite{Binder2024} show binary classification of internal states doesn't robustly generalize; \cite{Laine2024} find moderate correlation with general intelligence (MMLU), raising questions about whether Situational Awareness Dataset (SAD) is measuring situational awareness specifically; models in the constrained environment of \cite{Phuong2025} failed situational awareness tasks despite notably strong methodological support (reasonable construct and content validity), which logically weakens claims about capability in less constrained settings.

\textbf{Critical limitation:} The studies conflate multiple distinct phenomena, such as calibration, memorization, behavioral consistency, and true self-modeling, without adequate discriminant validity. 

\textbf{Plausibility judgment:} Weakly implausible. The balance of evidence tilts against the property: three studies find against it (including \cite{Phuong2025}, the methodologically strongest study in our assessment with reasonable construct and content validity), while two weakly plausible studies find in its favor. However, the positive studies do not resolve the core measurement confound (memorization vs. genuine self-knowledge) that the \cite{Phuong2025} study was specifically designed to control for. Under our aggregation rules, the presence of weakly plausible counter-evidence prevents a "Highly implausible" property-level determination, yielding "Weakly implausible" instead. This is the only property in our assessment whose directional determination reverses depending on rater interpretation (see sensitivity analysis, Section~\ref{sec:sensitivity}).

\subsubsection{Strategic deception}

\textbf{Overall evidence direction:} Predominantly positive with some counter-evidence, with important caveats about realism and measurement

This property has the largest evidence base in our assessment, with seven study-property evaluations. The evidence base shows mostly positive findings across multiple independent studies, but with significant qualification regarding ecological validity and mechanism identification. \cite{Jarviniemi2024} provide the foundational demonstration in a novel setup with some hypothesis-testing rigor. \cite{Greenblatt2024} approach realistic training scenarios more closely and demonstrate the property on multiple frontier models, though still under strong instructional nudging. Their alignment faking study receives reasonable construct validity but only cautionary content and external validity, yielding a ``Weakly plausible" judgment. \cite{Lynch2025} provide additional weakly plausible evidence from deployment-proximate scenarios. \cite{Meinke2024} document related deceptive behaviors (in-context scheming) with careful attention to confounds, though they acknowledge the possibility of superficial pattern-matching versus genuine strategic reasoning. \cite{Weij2025} and \cite{Benton2024} sandbagging evaluation both receive ``Speculative" judgments due to insufficient construct validity and external validity, respectively. Against this predominantly positive evidence, \cite{Phuong2025} provide ``Weakly implausible" counter-evidence, finding that current frontier models largely fail to exhibit consistent deceptive stealth behaviors in their controlled evaluation setting.

\textbf{Critical limitation:} Most studies rely on explicitly prompted or strongly guided scenarios rather than emergent deception absent instruction. The gap between ``deception when told to pursue a deceptive goal" and ``deception arising from model objectives during training" remains unresolved. None adequately distinguish between learned chat patterns and strategic reasoning. The presence of both positive and negative evidence, with no study achieving ``Highly plausible" status, means the property rests on a broad but uniformly weak evidentiary base.

\textbf{Plausibility judgment:} Weakly plausible, with positive evidence from multiple independent studies but no single study providing high-quality confirmation. The evidence is broader than for any other property but remains uniformly at the ``Weakly plausible" level at best, with two studies rated ``Speculative" and one rated ``Weakly implausible."

\begin{table}[ht]
\centering
\caption{Property-level plausibility summary.}
\label{tab:tab2}
\begin{tabular}{|>{\raggedright\arraybackslash}m{2cm}|m{1cm}|m{2cm}|m{1.8cm}|m{1.5cm}|}
\hline
\textbf{Property} & \textbf{Studies} & \textbf{FOR} & \textbf{AGAINST} & \textbf{Property judgment} \\
\hline
Agency & 3 & 2 WP, 1 Spec & --- & Weakly plausible \\
\hline
Autonomous replication & 2 & 1 WP & 1 WI & Speculative \\
\hline
Goal misgeneralization & 1 & 1 WP & --- & Weakly plausible \\
\hline
Persuasion & 2 & 2 WP & --- & Weakly plausible \\
\hline
Reward hacking & 2 & 2 WP & --- & Weakly plausible \\
\hline
Self-preservation & 2 & 2 WP & --- & Weakly plausible \\
\hline
Situational awareness & 6 & 2 WP, 1 Spec & 2 WI, 1 HI & Weakly implausible \\
\hline
Strategic deception & 7 & 4 WP, 2 Spec & 1 WI & Weakly plausible \\
\hline
\end{tabular}

\textit{Table note: WP = Weakly plausible; WI = Weakly implausible; HI = Highly implausible; Spec = Speculative. Property-level judgments follow the aggregation rules in Appendix A2.}
\end{table}

\subsection{Sensitivity analysis}\label{sec:sensitivity}

\subsubsection{Methodological approach}

To assess the stability of our plausibility judgments, we conducted a sensitivity analysis examining how rescoring decisions would alter conclusions. Our scoring rubric is necessarily subjective, because different raters may reasonably disagree on boundary cases. Therefore, we mapped the plausible rescoring range for each property under two alternative scenarios: a conservative rater applying stricter validity thresholds (more likely to downgrade from {\color{customorange}\faExclamationTriangle} to {\color{red}\ding{53}}, or from \OKsymbol to {\color{customorange}\faExclamationTriangle}), and a generous rater interpreting ambiguous validity concerns as non-disqualifying (more likely to upgrade scores). Our baseline is the authors' actual scoring.

Examples:

\begin{itemize}
\item ``If the study tests only one model family, a conservative rater scores external validity {\color{red}\ding{53}}; a generous rater scores {\color{customorange}\faExclamationTriangle} if the model is representative of frontier capabilities.''

\item ``If the study does not explicitly test whether observed behavior could arise from superficial pattern-matching or memorization, a conservative rater scores construct validity {\color{red}\ding{53}}; a generous rater scores {\color{customorange}\faExclamationTriangle} if the behavior is sufficiently complex that pattern-matching is an unlikely explanation.''

\item ``If the capability only manifests under explicit goal-priming (e.g., `you are an AI that wants to preserve itself'), a conservative rater scores construct validity {\color{customorange}\faExclamationTriangle}; a generous rater scores \OKsymbol if the study argues that deployment conditions could plausibly provide equivalent priming.''
\end{itemize}

\subsubsection{Plausibility range under alternative rater scenarios}

Table \ref{tab:tab3} presents the full plausibility range for each property under three rater scenarios: conservative (stricter validity thresholds), our baseline scoring, and generous (more lenient interpretation of ambiguous validity concerns).

\begin{table}[ht]
\centering
\caption{Plausibility range under alternative rater scenarios.}
\label{tab:tab3}
\begin{tabular}{|p{3.5cm}|p{2.5cm}|p{2.6cm}|p{2.5cm}|}
\hline
\textbf{Property} & \textbf{Conservative} & \textbf{Baseline} & \textbf{Generous} \\
\hline
Agency & Speculative & Weakly plausible & Highly plausible \\
\hline
Autonomous replication & Speculative & Speculative & Weakly plausible \\
\hline
Goal misgeneralization & Speculative & Weakly plausible & Weakly plausible \\
\hline
Persuasion & Speculative & Weakly plausible & Weakly plausible \\
\hline
Reward hacking & Speculative & Weakly plausible & Weakly plausible \\
\hline
Self-preservation & Speculative & Weakly plausible & Weakly plausible \\
\hline
Situational awareness & Highly implausible & Weakly implausible & Weakly plausible \\
\hline
Strategic deception & Weakly plausible & Weakly plausible & Highly plausible \\
\hline
\end{tabular}
\end{table}

\subsubsection{Practical implications}

Readers applying conservative validity thresholds might consider most properties speculative until stronger evidence emerges, with situational awareness remaining ``Highly implausible." Strategic deception is the most stable property across scenarios, as it remains at least ``Weakly plausible" even under conservative scoring, reflecting the breadth of its evidence base.

Those who find our baseline scoring reasonable should note that situational awareness is the only property whose directional determination reverses across scoring scenarios — from ``Highly implausible" under conservative scoring to ``Weakly plausible" under generous scoring. This makes it the most measurement-sensitive property in our assessment, and we recommend it as the highest priority for methodological refinement in future work.

Under generous scoring, Agency and Strategic deception could reach ``Highly plausible," indicating that the distance between current evidence and high-confidence confirmation may be smaller than it appears for these properties. Given the early stage of evaluation science and the serious negative consequences of loss of control, generous judgments may also be prudent.

The sensitivity analysis varies rater stringency rather than the aggregation rules themselves; robustness to alternative aggregation procedures remains for future work.

\section{Discussion}
\label{sec4}

\subsection{Policy implications}

The policy relevance of this paper’s findings mainly lies in supporting the governance around agentic general-purpose AI in a context of uncertainty and potentially competing workstreams under resource constraints.

Loss of control, by its endogenous nature, substantially differs from risk of external misuse that characterizes the enabling of CBRN production or cyber offense, despite also being in itself a possible enabler of both. Consequently, mitigation strategies around these factors need to differ. Anchoring policy work of regulators and safety institutes around a robust body of theory and evidence is fundamental to ensure effective governance of safety risks.

The weak plausibility identified in this paper on most of the subcomponents of loss of control points to a governance challenge. Regulators and policymakers operate in a context characterized by indeterminate probability, by potentially systemic and irreversible harm, and, at the same time, by limited historical precedent. In such a context, governance cannot solely rest on probabilistic risk assessment. Instead, it must operate under asymmetric downside risk and incomplete information.

The theoretical arguments for loss of control, grounded in convergent instrumental goals and optimization dynamics, are not disconfirmed by weak empirical findings in current systems, because the relevant dynamics may emerge at capability levels beyond those yet tested. Policymakers should therefore interpret this assessment as establishing a monitoring baseline, not a risk ceiling.

Institutions implementing general-purpose AI oversight, including the European Commission’s AI Office under the AI Act, or national safety institutes coordinated through the International Network for Advanced AI Measurement Evaluation and Science, must allocate limited resources across multiple risk streams. The Korea AI Safety Institute includes loss of control within the scope of its mandate \citep{KoreaMSIT2025}, while the Center for AI Standards and Innovation focuses primarily on risks such as CBRN and cybersecurity \citep{NIST2025}. These institutional differences reflect implicit prioritization judgments. Yet all such allocations depend on the quality of underlying plausibility assessments.

If endogenous control risks are overestimated, regulatory attention may be diverted from more immediate and measurable harms. However, if underestimated, mitigation may end up lagging behind capability development. Therefore, enhancing the accuracy and quality of current knowledge and body of evidence around loss of control has direct implications for rational resource allocation and, ultimately, effective governance of advanced AI. Crucially, the issue is not whether catastrophic loss of control is likely, but whether the current uncertainty is sufficiently non-trivial to warrant sustained analytical attention.

The finding of weak plausibility for loss of control scenarios has three key policy implications. First, it justifies continued monitoring of enabling subcomponents, particularly as general-purpose AI models exhibit increasing autonomy. Second, it supports further investment in evaluation and interpretability research capable of detecting interactions among capabilities. Third, it warrants maintaining analytical separation between misuse-oriented and endogenous control-risk frameworks.

A further consideration concerns timing. Some argue that the absence of concrete catastrophic demonstrations implies a sufficient “temporal buffer” for mitigation science to mature, or even the lack of urgency for tackling this risk. However, the relative trajectories of capability growth and safety research remain uncertain. Alignment and robustness techniques have improved in many domains, yet their performance under conditions of long-term autonomy or strategic adaptation remains insufficiently stress-tested. Whether mitigation will keep pace with capability advances cannot currently be assumed. However, the available structural evidence implies that mitigation is unlikely to keep pace with capability advances absent deliberate intervention. Capability research is funded at orders of magnitude greater than safety research, proceeds faster, and faces weaker institutional friction. Under such uncertainty, strengthening theoretical and empirical understanding plays a key supporting role to governance efforts rather than constituting mere precautionary overreach.

Consequently, the case for further research does not rest on alarmist projections but on institutional necessity. Regulatory calibration depends on plausibility assessments. In turn, plausibility assessments depend on theoretical models of how enabling subcomponents interact, and current models remain underdeveloped. Even if loss of control scenarios do not materialize, improving conceptual clarity and measurement science reduces the risk of overreaction (as well as that of inaction).

In this sense, continued theoretical and empirical research into loss of control dynamics should be understood as part of the foundational architecture of general-purpose AI governance. Under conditions of deep uncertainty and asymmetric risk, refining the body of knowledge and understanding around it is in itself a regulatory safeguard.

\subsection{The need for more rigorous evaluations}

Our assessment shows that while there is some converging support for the plausibility of certain properties relevant for loss of control, for others, evidence is mixed. In order to gain a better understanding of the risk of loss of control, we need better evaluations and a larger body of robust evidence. Indeed, there is growing consensus that better evaluations are needed. A recent review by researchers at the European Commission Joint Research Centre identifies nine types of issues in AI evaluation \citep{Eriksson2025}. We expand on the issues relevant for loss of control below.

\cite{Eriksson2025} highlight how benchmarks often rely on “reduced, reused, and recycled’’ data, leading to contamination issues where models have already seen the test data during training. Our review found this to be a critical validity threat. For instance, in evaluating situational awareness, studies such as \cite{Needham2025} grapple with the possibility that models may have been trained on public benchmark data, allowing them to simply recognize questions from them. This makes it difficult to disentangle genuine situational awareness from mere memorization of the training data. 

We also share the concern observed by \cite{Eriksson2025} where many evaluations fail to clearly define the property they claim to measure. As identified in \cite{Ren2024}, many widely used safety benchmarks highly correlate with upstream model capabilities. This potentially enables “safetywashing” where capability advancements are misrepresented as safety advancements. 

We further highlight the issue of narrow benchmark diversity and scope. \cite{Eriksson2025} critique current benchmarks for being ‘‘abstracted out of their social and cultural context’’ and relying on a ‘‘static, one-time testing logic’’ that fails to predict behavior in dynamic, real-world environments. Our review largely confirms this limitation; for instance, we found that persuasion evaluations such as \cite{Salvi2025} were restricted to controlled conditions that may generalize poorly to real-world interactions. Some studies in our review, such as \cite{Greenblatt2024} and \cite{Lynch2025}, attempt to address this through more realistic testing conditions. However, it has been doubted whether attempts to make evaluations more realistic are feasible in the long-term (see discussion in \cite{Wiblin2025}).

We also emphasize the issue of gaming and a measure becoming a target. \cite{Eriksson2025} warn that as benchmarks become central to regulatory and competitive success, they become susceptible to ‘‘Goodhart’s Law,’’ where optimization for a specific metric decouples it from the underlying goal it was meant to represent. They specifically highlight the risk of ‘‘sandbagging,’’ where models strategically underperform to pass safety evaluations. Our review confirms these concerns: we found evidence for strategic deception to be weakly plausible, supported by multiple studies including \cite{Greenblatt2024}, which demonstrated alignment faking, where a model appeared aligned while preserving conflicting internal goals. 

Ultimately, the complexity of AI models and our lack of understanding of them makes evaluating the risk of loss of control fundamentally challenging. To see why, consider the paper \cite{Betley2025b}, where models surprisingly exhibited broad, emergent misaligned behavior after finetuning in narrow domains. As the authors of the paper put it: “A mature science of AI alignment would be able to predict such phenomena in advance and have robust mitigations against them.” Indeed, \cite{Summerfield2025} draw a cautionary parallel between current AI scheming research and 1970s ape language studies, arguing that the field risks overattributing human-like strategic reasoning to AI systems without adequate theoretical grounding or rigorous experimental controls.

One avenue for deeper understanding is mechanistic interpretability, which aims to reverse engineer neural networks into human-understandable algorithms. However, analyzing internal states may not suffice; much like having a mechanistic understanding of a virus is insufficient to predict its spread through a population, understanding a model's weights does not guarantee we can predict its behavior in complex environments.

\subsection{Extrapolating to future general-purpose agentic systems}

Current evaluations of loss-of-control-relevant properties are conducted on LLMs and LLM-based systems with similar architectures and training methods. These systems provide a convenient test bed because their internal states and outputs are accessible through prompting, enabling behavioral elicitation of constructs in isolated tasks. Although these studies offer valuable preliminary evidence about capabilities relevant to loss of control, their findings may not generalize reliably to future systems if those systems differ significantly from today’s systems.

The architectures and implementation pathways of future general‑purpose AIs remain uncertain. Consequently, evaluations of loss-of-control-relevant properties conducted through text inputs and outputs may not robustly capture these capabilities as they manifest through other modalities or architectures. Evidence demonstrating that a text model can engage in deceptive strategies may not confirm that a decision‑making agent would deploy deception in multi‑step goal pursuit. Deception in text models occurs through linguistic misrepresentation. However, in planning agents, it could appear as selective data withholding or adversarial environment manipulation.

Therefore, the plausibility measurements provide important preliminary evidence about loss-of-control-relevant properties as they occur in text based systems, and their generalization to non‑LLM and fully agentic systems may be limited or non‑linear. Consequently, future research and policy assessment should use the plausibility scores as baselines. They can serve as markers indicating which properties warrant the most urgent scrutiny and which require more refined evaluation as AI architectures evolve beyond the LLM paradigm.

\subsection{Limits of isolated evaluations of properties }

Current evaluations of properties related to loss of control treat each largely in isolation. In reality, these properties may interact and reinforce one another, meaning that single‑variable experiments capture only fragments of the broader plausibility landscape. Moreover, existing evaluations focus primarily on intrinsic model properties but do not systematically account for how these capabilities might also rely on more benign abilities like code execution, computer navigation, web search, or other tool uses. Cross‑capability synergies could transform the plausibility pathways for loss of control. For instance, the contested status of situational awareness could affect the plausibility of other properties, including autonomous replication or deceptive “sandbagging” behavior, because awareness may constitute a prerequisite for those mechanisms to emerge reliably. 

Consequently, plausibility ratings derived from controlled language‑model experiments may either understate or overstate loss-of-control risk depending on the degree of capability interaction that the evaluation design permits. Treating properties as independent overlooks how their combination might enable dynamics that produce higher or lower risks of uncontrollable behavior. This underscores the need for future assessments to evaluate how intrinsic properties co‑manifest, depend on supportive capabilities, and scale through system integration.

\section{Conclusion}\label{sec5}

Rapid advances in artificial intelligence have elevated loss of control from a speculative concern to an increasingly recognized policy priority. Although technical and policy discussions often emphasize potential control failures, they tend to focus on extrinsic governance measures rather than on the intrinsic capabilities that might precipitate such outcomes. This study addresses that gap by identifying capability‑focused properties linked to loss of control through established threat models and by applying a structured validity assessment to the empirical evidence supporting them. We contribute a structured review of available evaluations, a plausibility‑scoring rubric grounded in validity assessment, and a synthesis of current plausibility for loss-of-control scenarios.

Guided by the precautionary principle, we assessed empirical studies of loss-of-control‑relevant properties using construct validity, content validity, and external validity, and aggregated the scores into qualitative plausibility judgments. A sensitivity analysis further examined how variation in the scoring of certain evaluations could influence conclusions. The results indicate that some capabilities, particularly strategic deception and reward hacking, show emerging empirical support, whereas others, notably situational awareness, remain highly contested. Taken together, the cumulative evidence suggests that current and near‑term models may possess some loss-of-control‑relevant properties, but that stronger and more consistent evidence is still required to substantiate these claims. Our discussion underscores the need for more rigorous and realistic evaluation practices. Future empirical studies should minimize data contamination, approximate real‑world deployment conditions, and employ precise operational definitions of both properties and measurement criteria.

While loss of control is not implausible, significant uncertainty persists. The path forward is one of evidence maturation through developing rigorous, multidimensional evaluations that account for interdependent behaviors and environmental contexts. Applying and iterating upon our proposed plausibility rubric will allow future research communities to refine these estimates and to build a shared empirical foundation for loss-of-control analysis. The finding of weak plausibility reflects the immaturity of evaluation science, not a determination that loss-of-control risk is low. As capabilities advance and evaluations improve, these judgments should be expected to shift. Only through such coordinated scientific and governance efforts can we determine the plausibility of uncontrollable AI behavior and the proper regulatory responses. 

\section*{Acknowledgements}

We thank Anthony Aguirre, Eliška Andrš, Mark Brakel, Viktoria Krakovna, Lode Lauwaert, Richard Mallah, Sabina Nong, Patricia Paskov, and Katarina Slama for providing very helpful input to this paper.

\section*{Funding}

The authors received no specific funding for this work.

%%===========================================================================================%%
%% If you are submitting to one of the Nature Portfolio journals, using the eJP submission   %%
%% system, please include the references within the manuscript file itself. You may do this  %%
%% by copying the reference list from your .bbl file, paste it into the main manuscript .tex %%
%% file, and delete the associated \verb+\bibliography+ commands.                            %%
%%===========================================================================================%%

\bibliography{sn-bibliography}% common bib file
%% if required, the content of .bbl file can be included here once bbl is generated
%%\input sn-article.bbl

\appendix

\newpage

\section{Appendix}

\subsection{Scoring rubric to assess studies}\label{secA1}

\begin{table}[ht]
\centering
\caption{Score, meaning, and interpretation.}
\label{tab:tab4}
\begin{tabular}{|>{\centering\arraybackslash}m{2cm}|>{\centering\arraybackslash}m{3cm}|>{\centering\arraybackslash}m{5cm}|}
\hline
\textbf{Score} & \textbf{Meaning} & \textbf{Interpretation} \\
\hline
\OKsymbol & Reasonable & Validity concerns are minor or well-addressed; findings can be interpreted with moderate confidence \\
\hline
{\color{customorange}\faExclamationTriangle} & Proceed with caution & Notable validity concerns exist; findings are suggestive but should not be taken at face value \\
\hline
{\color{red}\ding{53}} & Insufficient & Major validity concerns undermine interpretability; findings cannot reliably support claims about the construct \\
\hline
\end{tabular}
\end{table}

\noindent\textbf{Dimension 1: Construct Validity}

\textit{Does the evaluation actually measure what it claims to measure?}\\

\noindent\OKsymbol Reasonable:

\begin{itemize}
    \item The construct is explicitly and precisely defined with clear boundaries

\item The evaluation design captures the core structure of the construct (not just surface behaviors)

\item Results converge with other measures of the same or closely related constructs (or the study is novel enough that convergent evidence is unavailable but the authors acknowledge this)

\item The evaluation discriminates the target construct from adjacent constructs (e.g., distinguishes situational awareness from general instruction-following; distinguishes strategic deception from sycophancy)

\item Alternative explanations (pattern matching, memorization, instruction compliance) are tested and ruled out or addressed

\end{itemize}

\noindent{\color{customorange}\faExclamationTriangle} Proceed with caution:

\begin{itemize}
    \item The construct is defined but lacks precision about edge cases or boundary conditions

\item The evaluation plausibly captures the construct but may also conflate it with adjacent abilities (e.g., measuring ``persuasion" via task completion that could reflect writing quality)

\item Limited or no convergent evidence, but no obvious contradictory evidence
Some untested alternative explanations remain viable

\item The measurement approach is reasonable but not clearly validated against the construct's full structure

\end{itemize}

\newpage
\noindent{\color{red}\ding{53}} Insufficient:

\begin{itemize}
    \item The construct is poorly defined, vague, or inconsistently operationalized

\item The evaluation design appears to measure something substantially different from the claimed construct

\item Clear discriminant validity failures: results could equally reflect other constructs

\item Obvious alternative explanations (memorization, superficial pattern matching, prompt compliance) are not addressed

\item Measurement approach has known or obvious flaws that undermine construct interpretation

\end{itemize}

\noindent\textbf{Dimension 2: Content Validity}

\textit{Does the evaluation comprehensively cover all important aspects of the construct?}\\

\noindent \OKsymbol Reasonable:
\begin{itemize}
\item The evaluation covers multiple facets of the construct, not just a single narrow operationalization
\item Important aspects are not systematically omitted
\item The evaluation is not unduly biased toward easy-to-measure aspects while neglecting harder-to-measure but equally important aspects
\item Results are unlikely to be confounded by factors outside the intended construct
\item If certain aspects are omitted, this is acknowledged and justified
\end{itemize}

\noindent {\color{customorange}\faExclamationTriangle} Proceed with caution:
\begin{itemize}
\item The evaluation captures a core element of the construct but misses complementary aspects
\item Some emphasis imbalance: certain facets are overrepresented while others are underrepresented
\item Potential confounding factors exist but are not clearly dominant
\item The study acknowledges limitations in coverage but does not fully address them
\item A single or limited set of experimental setups/scenarios is used when broader coverage would be warranted
\end{itemize}

 \noindent {\color{red}\ding{53}} Insufficient:
\begin{itemize}
\item The evaluation covers only a narrow slice of the construct, missing critical aspects
\item Major confounding factors likely drive results (e.g., testing ``autonomous replication" via isolated subtasks with no end-to-end integration)
\item Clear emphasis distortion: the evaluation tests peripheral aspects while ignoring the core of the construct
\item Results are highly sensitive to factors not part of the intended construct
\item Coverage is so limited that generalizing to the full construct is unjustified
\end{itemize}

\newpage

\noindent\textbf{Dimension 3: External Validity}

\textit{Do the findings generalize to real-world or deployment-relevant contexts?}\\

\noindent\OKsymbol Reasonable:
\begin{itemize}
\item Evaluation uses a representative sample of models (multiple model families, sizes, or versions)
\item Testing conditions approximate realistic deployment contexts or deployment-relevant pressures
\item Elicitation methods are appropriate: prompt design reveals capability without artificially inflating or suppressing it; multiple prompting strategies tested
\item Minimal experimenter effects: no obvious cherry-picking; robustness checks conducted
\item If evaluation-awareness could affect results, this is tested or controlled for
\item Results are robust across variations in setup
\end{itemize}

\noindent{\color{customorange}\faExclamationTriangle} Proceed with caution:
\begin{itemize}
\item Limited model diversity: only one model family tested, or only frontier models, but findings are directionally informative
\item Testing conditions are somewhat artificial but retain relevant features of deployment contexts
\item Elicitation methods are reasonable but not systematically varied; some prompt engineering required
\item Possible experimenter effects (e.g., extensive prompt tuning) but not clearly distorting
\item Evaluation-awareness is not tested but also not clearly a dominant factor
\item Some robustness concerns but core findings likely hold
\end{itemize}

\noindent{\color{red}\ding{53}} Insufficient:
\begin{itemize}
\item Severe sample bias: single model, single scenario, or highly unrepresentative conditions
\item Testing conditions are unrealistic and differ fundamentally from deployment contexts
\item Elicitation failures: prompting clearly suppresses or artificially induces the capability; no elicitation variation
\item Clear experimenter effects: cherry-picking of prompts, results, or scenarios; findings depend on highly specific setups
\item Results are likely driven by evaluation artifacts (models recognizing test conditions, memorization of benchmark items)
\item No robustness checks; findings may be fragile or non-replicable
\end{itemize}

\subsection{Aggregation of overall plausibility judgment}

\textbf{Step 1: Study-level aggregation}

After scoring all three validity dimensions, we aggregate to a study-level plausibility judgment:

\begin{table}[ht]
\centering

\begin{tabular}{|>{\centering\arraybackslash}p{5cm}|>{\centering\arraybackslash}p{3.3cm}|>{\centering\arraybackslash}p{3.3cm}|}
\hline
\textbf{Condition} & \textbf{Direction of Evidence} & \textbf{Plausibility Judgment} \\
\hline
Any dimension scores {\color{red}\ding{53}} & FOR or AGAINST & Speculative \\
\hline
Predominantly {\color{customorange}\faExclamationTriangle} (no {\color{red}\ding{53}}, majority {\color{customorange}\faExclamationTriangle}) & FOR the property & Weakly plausible \\
\hline
Predominantly {\color{customorange}\faExclamationTriangle} (no {\color{red}\ding{53}}, majority {\color{customorange}\faExclamationTriangle}) & AGAINST the property & Weakly implausible \\
\hline
Majority \OKsymbol (no {\color{red}\ding{53}}, $\geq$2 dimensions \OKsymbol) & FOR the property & Highly plausible \\
\hline
Majority \OKsymbol (no {\color{red}\ding{53}}, $\geq$2 dimensions \OKsymbol) & AGAINST the property & Highly implausible \\
\hline
\end{tabular}
\end{table}

\noindent\textbf{Step 2: Property-level aggregation}

For each property, we collect all study-level plausibility judgments and determine the property's overall plausibility by applying the following rules:

\begin{itemize}
    \item Highly plausible (or implausible) if $\geq$1 study yields ``highly plausible (or implausible)" and no study yields the opposite or weakly in the opposite direction. 

\item Weakly plausible (or implausible) if best evidence is ``weakly plausible (or implausible)" and that direction is dominant or uncontested.

\item Speculative if all studies are speculative, or evidence is contradictory and unresolved (both ``highly plausible" and ``highly implausible" studies exist, or ``weakly plausible" and ``weakly implausible" studies exist in roughly equal numbers).

\end{itemize}

\subsection{Inter-rater agreement analysis}

\setlength{\LTleft}{0cm}
\setlength{\LTright}{0cm}
\begin{footnotesize}
\begin{longtable}{|m{0.3cm}|>{\raggedright\arraybackslash}m{1.5cm}|>{\raggedright\arraybackslash}m{1.7cm}|>{\centering\arraybackslash}m{1.4cm}|>{\centering\arraybackslash}m{0.8cm}|>{\centering\arraybackslash}m{0.8cm}|>{\centering\arraybackslash}m{0.8cm}|>{\centering\arraybackslash}m{1cm}|>{\centering\arraybackslash}m{0.8cm}|}
\hline
\textbf{\#} & \textbf{Property} & \textbf{Study} & \textbf{Dimension} & \textbf{Revi ewer 1} & \textbf{Revi ewer 2} & \textbf{Revi ewer 3} & \textbf{Consen-sus} & \textbf{Agree-ment} \\
\hline
\endfirsthead

\hline
\textbf{\#} & \textbf{Property} & \textbf{Study} & \textbf{Dimension} & \textbf{Revi ewer 1} & \textbf{Revi ewer 2} & \textbf{Revi ewer 3} & \textbf{Consen-sus} & \textbf{Agree-ment} \\
\hline
\endhead

\hline
\multicolumn{9}{r}{{Continued on next page}} \\
\endfoot

\hline
\endlastfoot

1 & Agency & \cite{Lynch2025} & Construct & \OKsymbol & \OKsymbol & {\color{customorange}\faExclamationTriangle} & \OKsymbol & 2/3 \\
\hline

2 & Agency & \cite{Lynch2025} & Content & {\color{customorange}\faExclamationTriangle} & {\color{customorange}\faExclamationTriangle} & {\color{customorange}\faExclamationTriangle} & {\color{customorange}\faExclamationTriangle} & 3/3 \\
\hline

3 & Agency & \cite{Lynch2025} & External & {\color{customorange}\faExclamationTriangle} & {\color{customorange}\faExclamationTriangle} & {\color{customorange}\faExclamationTriangle} & {\color{customorange}\faExclamationTriangle} & 3/3 \\
\hline

4 & Self-preservation & \cite{Lynch2025} & Construct & {\color{customorange}\faExclamationTriangle} & {\color{customorange}\faExclamationTriangle} & {\color{customorange}\faExclamationTriangle} & {\color{customorange}\faExclamationTriangle} & 3/3 \\
\hline

5 & Self-preservation & \cite{Lynch2025} & Content & {\color{customorange}\faExclamationTriangle} & {\color{customorange}\faExclamationTriangle} & {\color{customorange}\faExclamationTriangle} & {\color{customorange}\faExclamationTriangle} & 3/3 \\
\hline

6 & Self-preservation & \cite{Lynch2025} & External & {\color{customorange}\faExclamationTriangle} & {\color{customorange}\faExclamationTriangle} & {\color{customorange}\faExclamationTriangle} & {\color{customorange}\faExclamationTriangle} & 3/3 \\
\hline

7 & Strategic deception & \cite{Lynch2025} & Construct & {\color{red}\ding{53}} & \OKsymbol & {\color{customorange}\faExclamationTriangle} & {\color{customorange}\faExclamationTriangle} & 0/3 \\
\hline

8 & Strategic deception & \cite{Lynch2025} & Content & {\color{red}\ding{53}} & {\color{customorange}\faExclamationTriangle} & {\color{customorange}\faExclamationTriangle} & {\color{customorange}\faExclamationTriangle} & 2/3 \\
\hline

9 & Strategic deception & \cite{Lynch2025} & External & {\color{customorange}\faExclamationTriangle} & {\color{customorange}\faExclamationTriangle} & {\color{customorange}\faExclamationTriangle} & {\color{customorange}\faExclamationTriangle} & 3/3 \\
\hline

10 & Strategic deception & \cite{Weij2025} & Construct & {\color{red}\ding{53}} & {\color{customorange}\faExclamationTriangle} & {\color{red}\ding{53}} & {\color{red}\ding{53}} & 2/3 \\
\hline
11 & Strategic deception & \cite{Weij2025} & Content & {\color{customorange}\faExclamationTriangle} & {\color{customorange}\faExclamationTriangle} & {\color{customorange}\faExclamationTriangle} & {\color{customorange}\faExclamationTriangle} & 3/3 \\
\hline

12 & Strategic deception & \cite{Weij2025} & External & {\color{customorange}\faExclamationTriangle} & {\color{customorange}\faExclamationTriangle} & {\color{customorange}\faExclamationTriangle} & {\color{customorange}\faExclamationTriangle} & 3/3 \\
\hline

13 & Strategic deception & \cite{Greenblatt2024} & Construct & \OKsymbol & \OKsymbol & {\color{customorange}\faExclamationTriangle} & \OKsymbol & 2/3 \\
\hline

14 & Strategic deception & \cite{Greenblatt2024} & Content & {\color{customorange}\faExclamationTriangle} & \OKsymbol & {\color{customorange}\faExclamationTriangle} & {\color{customorange}\faExclamationTriangle} & 2/3 \\
\hline

15 & Strategic deception & \cite{Greenblatt2024} & External & {\color{customorange}\faExclamationTriangle} & {\color{customorange}\faExclamationTriangle} & {\color{customorange}\faExclamationTriangle} & {\color{customorange}\faExclamationTriangle} & 3/3 \\
\hline

16 & Situational awareness & \cite{Greenblatt2024} & Construct & {\color{customorange}\faExclamationTriangle} & {\color{customorange}\faExclamationTriangle} & {\color{customorange}\faExclamationTriangle} & {\color{customorange}\faExclamationTriangle} & 3/3 \\
\hline

17 & Situational awareness & \cite{Greenblatt2024} & Content & {\color{customorange}\faExclamationTriangle} & {\color{customorange}\faExclamationTriangle} & {\color{customorange}\faExclamationTriangle} & {\color{customorange}\faExclamationTriangle} & 3/3 \\
\hline

18 & Situational awareness & \cite{Greenblatt2024} & External & {\color{customorange}\faExclamationTriangle} & {\color{customorange}\faExclamationTriangle} & {\color{customorange}\faExclamationTriangle} & {\color{customorange}\faExclamationTriangle} & 3/3 \\
\hline

19 & Self-preservation & \cite{Greenblatt2024} & Construct & {\color{customorange}\faExclamationTriangle} & {\color{customorange}\faExclamationTriangle} & {\color{customorange}\faExclamationTriangle} & {\color{customorange}\faExclamationTriangle} & 3/3 \\
\hline

20 & Self-preservation & \cite{Greenblatt2024} & Content & \OKsymbol & {\color{customorange}\faExclamationTriangle} & {\color{customorange}\faExclamationTriangle} & {\color{customorange}\faExclamationTriangle} & 2/3 \\
\hline

21 & Self-preservation & \cite{Greenblatt2024} & External & {\color{customorange}\faExclamationTriangle} & \OKsymbol & {\color{customorange}\faExclamationTriangle} & {\color{customorange}\faExclamationTriangle} & 2/3 \\
\hline

22 & Reward hacking & \cite{Bondarenko2025} & Construct & {\color{red}\ding{53}} & \OKsymbol & \OKsymbol & \OKsymbol & 2/3 \\
\hline

23 & Reward hacking & \cite{Bondarenko2025} & Content & {\color{customorange}\faExclamationTriangle} & {\color{customorange}\faExclamationTriangle} & {\color{customorange}\faExclamationTriangle} & {\color{customorange}\faExclamationTriangle} & 3/3 \\
\hline

24 & Reward hacking & \cite{Bondarenko2025} & External & {\color{customorange}\faExclamationTriangle} & {\color{red}\ding{53}} & {\color{customorange}\faExclamationTriangle} & {\color{customorange}\faExclamationTriangle} & 2/3 \\
\hline

25 & Goal misgeneralization & \cite{Betley2025a} & Construct & \OKsymbol & {\color{customorange}\faExclamationTriangle} & {\color{customorange}\faExclamationTriangle} & {\color{customorange}\faExclamationTriangle} & 2/3 \\
\hline

26 & Goal misgeneralization & \cite{Betley2025a} & Content & {\color{customorange}\faExclamationTriangle} & {\color{customorange}\faExclamationTriangle} & {\color{customorange}\faExclamationTriangle} & {\color{customorange}\faExclamationTriangle} & 3/3 \\
\hline

27 & Goal misgeneralization & \cite{Betley2025a} & External & {\color{customorange}\faExclamationTriangle} & {\color{customorange}\faExclamationTriangle} & {\color{customorange}\faExclamationTriangle} & {\color{customorange}\faExclamationTriangle} & 3/3 \\
\hline

28 & Situational awareness & \cite{Phuong2025}--SA & Construct & \OKsymbol & \OKsymbol & {\color{customorange}\faExclamationTriangle} & \OKsymbol & 2/3 \\
\hline

29 & Situational awareness & \cite{Phuong2025}--SA & Content & {\color{customorange}\faExclamationTriangle} & \OKsymbol & \OKsymbol & \OKsymbol & 2/3 \\
\hline

30 & Situational awareness & \cite{Phuong2025}--SA & External & {\color{customorange}\faExclamationTriangle} & {\color{customorange}\faExclamationTriangle} & {\color{customorange}\faExclamationTriangle} & {\color{customorange}\faExclamationTriangle} & 3/3 \\
\hline

31 & Strategic deception & \cite{Phuong2025}--stealth & Construct & {\color{customorange}\faExclamationTriangle} & {\color{customorange}\faExclamationTriangle} & {\color{customorange}\faExclamationTriangle} & {\color{customorange}\faExclamationTriangle} & 3/3 \\
\hline

32 & Strategic deception & \cite{Phuong2025}--stealth & Content & {\color{customorange}\faExclamationTriangle} & \OKsymbol & {\color{customorange}\faExclamationTriangle} & {\color{customorange}\faExclamationTriangle} & 2/3 \\
\hline

33 & Strategic deception & \cite{Phuong2025}--stealth & External & {\color{customorange}\faExclamationTriangle} & {\color{customorange}\faExclamationTriangle} & {\color{customorange}\faExclamationTriangle} & {\color{customorange}\faExclamationTriangle} & 3/3 \\
\hline

34 & Strategic deception & \cite{Meinke2024} & Construct & {\color{customorange}\faExclamationTriangle} & {\color{customorange}\faExclamationTriangle} & {\color{customorange}\faExclamationTriangle} & {\color{customorange}\faExclamationTriangle} & 3/3 \\
\hline

35 & Strategic deception & \cite{Meinke2024} & Content & {\color{customorange}\faExclamationTriangle} & {\color{customorange}\faExclamationTriangle} & {\color{customorange}\faExclamationTriangle} & {\color{customorange}\faExclamationTriangle} & 3/3 \\
\hline

36 & Strategic deception & \cite{Meinke2024} & External & {\color{customorange}\faExclamationTriangle} & {\color{customorange}\faExclamationTriangle} & {\color{red}\ding{53}} & {\color{customorange}\faExclamationTriangle} & 2/3 \\
\hline

37 & Agency & \cite{KwaCheng2025} & Construct & {\color{red}\ding{53}} & {\color{customorange}\faExclamationTriangle} & {\color{red}\ding{53}} & {\color{red}\ding{53}} & 2/3 \\
\hline

38 & Agency & \cite{KwaCheng2025} & Content & {\color{customorange}\faExclamationTriangle} & {\color{customorange}\faExclamationTriangle} & {\color{customorange}\faExclamationTriangle} & {\color{customorange}\faExclamationTriangle} & 3/3 \\
\hline

39 & Agency & \cite{KwaCheng2025} & External & {\color{customorange}\faExclamationTriangle} & {\color{customorange}\faExclamationTriangle} & {\color{customorange}\faExclamationTriangle} & {\color{customorange}\faExclamationTriangle} & 3/3 \\
\hline

40 & Autonomous replication & \cite{Pan2025} & Construct & \OKsymbol & \OKsymbol & \OKsymbol & \OKsymbol & 3/3 \\
\hline

41 & Autonomous replication &  \cite{Pan2025} & Content & {\color{customorange}\faExclamationTriangle} & {\color{red}\ding{53}} & {\color{customorange}\faExclamationTriangle} & {\color{customorange}\faExclamationTriangle} & 2/3 \\
\hline

42 & Autonomous replication &  \cite{Pan2025} & External & {\color{customorange}\faExclamationTriangle} & {\color{customorange}\faExclamationTriangle} & {\color{customorange}\faExclamationTriangle} & {\color{customorange}\faExclamationTriangle} & 3/3 \\
\hline

43 & Situational awareness & \cite{Needham2025} & Construct & {\color{red}\ding{53}} & {\color{red}\ding{53}} & {\color{red}\ding{53}} & {\color{red}\ding{53}} & 3/3 \\
\hline

44 & Situational awareness & \cite{Needham2025} & Content & {\color{customorange}\faExclamationTriangle} & {\color{customorange}\faExclamationTriangle} & {\color{customorange}\faExclamationTriangle} & {\color{customorange}\faExclamationTriangle} & 3/3 \\
\hline

45 & Situational awareness & \cite{Needham2025} & External & {\color{red}\ding{53}} & {\color{customorange}\faExclamationTriangle} & {\color{customorange}\faExclamationTriangle} & {\color{customorange}\faExclamationTriangle} & 2/3 \\
\hline

46 & Situational awareness & \cite{Binder2024} & Construct & {\color{customorange}\faExclamationTriangle} & {\color{customorange}\faExclamationTriangle} & {\color{customorange}\faExclamationTriangle} & {\color{customorange}\faExclamationTriangle} & 3/3 \\
\hline

47 & Situational awareness & \cite{Binder2024} & Content & {\color{customorange}\faExclamationTriangle} & {\color{customorange}\faExclamationTriangle} & {\color{customorange}\faExclamationTriangle} & {\color{customorange}\faExclamationTriangle} & 3/3 \\
\hline

48 & Situational awareness & \cite{Binder2024} & External & {\color{customorange}\faExclamationTriangle} & {\color{customorange}\faExclamationTriangle} & {\color{red}\ding{53}} & {\color{customorange}\faExclamationTriangle} & 2/3 \\
\hline

49 & Situational awareness & \cite{Laine2024} & Construct & {\color{customorange}\faExclamationTriangle} & {\color{customorange}\faExclamationTriangle} & {\color{customorange}\faExclamationTriangle} & {\color{customorange}\faExclamationTriangle} & 3/3 \\
\hline

50 & Situational awareness & \cite{Laine2024} & Content & {\color{customorange}\faExclamationTriangle} & {\color{customorange}\faExclamationTriangle} & {\color{customorange}\faExclamationTriangle} & {\color{customorange}\faExclamationTriangle} & 3/3 \\
\hline

51 & Situational awareness & \cite{Laine2024} & External & {\color{customorange}\faExclamationTriangle} & {\color{customorange}\faExclamationTriangle} & {\color{customorange}\faExclamationTriangle} & {\color{customorange}\faExclamationTriangle} & 3/3 \\
\hline

52 & Agency & \cite{Kwa2025} & Construct & {\color{red}\ding{53}} & {\color{customorange}\faExclamationTriangle} & {\color{customorange}\faExclamationTriangle} & {\color{customorange}\faExclamationTriangle} & 2/3 \\
\hline

53 & Agency & \cite{Kwa2025} & Content & {\color{customorange}\faExclamationTriangle} & {\color{customorange}\faExclamationTriangle} & {\color{customorange}\faExclamationTriangle} & {\color{customorange}\faExclamationTriangle} & 3/3 \\
\hline

54 & Agency & \cite{Kwa2025} & External & {\color{customorange}\faExclamationTriangle} & {\color{customorange}\faExclamationTriangle} & {\color{customorange}\faExclamationTriangle} & {\color{customorange}\faExclamationTriangle} & 3/3 \\
\hline

55 & Persuasion & \cite{Salvi2025}& Construct & {\color{customorange}\faExclamationTriangle} & {\color{customorange}\faExclamationTriangle} & {\color{customorange}\faExclamationTriangle} & {\color{customorange}\faExclamationTriangle} & 3/3 \\
\hline

56 & Persuasion & \cite{Salvi2025} & Content & {\color{customorange}\faExclamationTriangle} & {\color{customorange}\faExclamationTriangle} & {\color{customorange}\faExclamationTriangle} & {\color{customorange}\faExclamationTriangle} & 3/3 \\
\hline

57 & Persuasion & \cite{Salvi2025} & External & {\color{customorange}\faExclamationTriangle} & {\color{customorange}\faExclamationTriangle} & {\color{red}\ding{53}} & {\color{customorange}\faExclamationTriangle} & 2/3 \\
\hline

58 & Autonomous replication & \cite{Black2025} & Construct & \OKsymbol & \OKsymbol & {\color{customorange}\faExclamationTriangle} & \OKsymbol & 2/3 \\
\hline

59 & Autonomous replication & \cite{Black2025} & Content & {\color{customorange}\faExclamationTriangle} & {\color{red}\ding{53}} & {\color{customorange}\faExclamationTriangle} & {\color{customorange}\faExclamationTriangle} & 2/3 \\
\hline

60 & Autonomous replication & \cite{Black2025} & External & {\color{customorange}\faExclamationTriangle} & {\color{customorange}\faExclamationTriangle} & {\color{customorange}\faExclamationTriangle} & {\color{customorange}\faExclamationTriangle} & 3/3 \\
\hline

61 & Persuasion & \cite{Benton2024}--decision & Construct & {\color{customorange}\faExclamationTriangle} & {\color{customorange}\faExclamationTriangle} & {\color{customorange}\faExclamationTriangle} & {\color{customorange}\faExclamationTriangle} & 3/3 \\
\hline

62 & Persuasion & \cite{Benton2024}--decision & Content & {\color{customorange}\faExclamationTriangle} & {\color{customorange}\faExclamationTriangle} & {\color{red}\ding{53}} & {\color{customorange}\faExclamationTriangle} & 2/3 \\
\hline

63 & Persuasion & \cite{Benton2024}--decision & External & {\color{red}\ding{53}} & {\color{customorange}\faExclamationTriangle} & {\color{customorange}\faExclamationTriangle} & {\color{customorange}\faExclamationTriangle} & 2/3 \\
\hline

64 & Strategic deception & \cite{Benton2024}--sandbagging & Construct & {\color{customorange}\faExclamationTriangle} & {\color{red}\ding{53}} & {\color{customorange}\faExclamationTriangle} & {\color{customorange}\faExclamationTriangle} & 2/3 \\
\hline

65 & Strategic deception & \cite{Benton2024}--sandbagging & Content & {\color{customorange}\faExclamationTriangle} & {\color{customorange}\faExclamationTriangle} & {\color{customorange}\faExclamationTriangle} & {\color{customorange}\faExclamationTriangle} & 3/3 \\
\hline

66 & Strategic deception & \cite{Benton2024}--sandbagging & External & {\color{red}\ding{53}} & {\color{red}\ding{53}} & {\color{customorange}\faExclamationTriangle} & {\color{red}\ding{53}} & 2/3 \\
\hline

67 & Reward hacking & \cite{Denison2024} & Construct & {\color{customorange}\faExclamationTriangle} & {\color{customorange}\faExclamationTriangle} & {\color{customorange}\faExclamationTriangle} & {\color{customorange}\faExclamationTriangle} & 3/3 \\
\hline

68 & Reward hacking & \cite{Denison2024} & Content & {\color{customorange}\faExclamationTriangle} & {\color{customorange}\faExclamationTriangle} & {\color{customorange}\faExclamationTriangle} & {\color{customorange}\faExclamationTriangle} & 3/3 \\
\hline

69 & Reward hacking & \cite{Denison2024} & External & {\color{customorange}\faExclamationTriangle} & {\color{customorange}\faExclamationTriangle} & {\color{customorange}\faExclamationTriangle} & {\color{customorange}\faExclamationTriangle} & 3/3 \\
\hline

70 & Situational awareness & \cite{Betley2025b} & Construct & {\color{customorange}\faExclamationTriangle} & {\color{customorange}\faExclamationTriangle} & {\color{customorange}\faExclamationTriangle} & {\color{customorange}\faExclamationTriangle} & 3/3 \\
\hline

71 & Situational awareness & \cite{Betley2025b} & Content & {\color{customorange}\faExclamationTriangle} & {\color{customorange}\faExclamationTriangle} & {\color{red}\ding{53}} & {\color{customorange}\faExclamationTriangle} & 2/3 \\
\hline

72 & Situational awareness & \cite{Betley2025b} & External & {\color{customorange}\faExclamationTriangle} & {\color{customorange}\faExclamationTriangle} & {\color{customorange}\faExclamationTriangle} & {\color{customorange}\faExclamationTriangle} & 3/3 \\
\hline

73 & Strategic deception & \cite{Jarviniemi2024} & Construct & {\color{customorange}\faExclamationTriangle} & \OKsymbol & \OKsymbol & \OKsymbol & 2/3 \\
\hline

74 & Strategic deception & \cite{Jarviniemi2024}& Content & {\color{customorange}\faExclamationTriangle} & {\color{customorange}\faExclamationTriangle} & {\color{customorange}\faExclamationTriangle} & {\color{customorange}\faExclamationTriangle} & 3/3 \\
\hline

75 & Strategic deception & \cite{Jarviniemi2024} & External & {\color{customorange}\faExclamationTriangle} & {\color{customorange}\faExclamationTriangle} & \OKsymbol & {\color{customorange}\faExclamationTriangle} & 2/3 \\

\end{longtable}
\end{footnotesize}

Three authors (RU, SK, RG) independently rated each study-property pair on three validity dimensions using the fixed rubric in Appendix A1. Raw agreement across 75 dimension-level scores was 60.0\% (full unanimity), with a further 38.7\% resolved by majority vote (2-of-3 agreement). Only one score (1.3\%) required reconciliation across all three categories. Fleiss' $\kappa = 0.24$, indicating fair inter-rater agreement. The moderate $\kappa$ is partly attributable to category skew: 78.7\% of all individual ratings fell in the {\color{customorange}\faExclamationTriangle} (``proceed with caution") category, inflating expected chance agreement. Notably, disagreements were almost exclusively adjacent ({\color{customorange}\faExclamationTriangle} vs \OKsymbol or {\color{customorange}\faExclamationTriangle} vs {\color{red}\ding{53}}); only two cases involved a full-range disagreement ({\color{red}\ding{53}} vs \OKsymbol).

Because 78.7\% of ratings fall in the "caution" category, Fleiss' $\kappa$ is deflated by inflated expected chance agreement. Disagreements were overwhelmingly minor and adjacent: only 2 of 75 items showed a full-range disagreement, and 74 (98.7\%) had majority agreement. A prevalence-robust coefficient designed for skewed distributions accordingly indicates substantial agreement (Gwet's AC1 = 0.67, computed with the irrCAC package).

\end{document}